\documentclass[preprint]{aastex62}

\graphicspath{{./}{figures/}}

\submitjournal{PSJ}

\shorttitle{Obliquity excitations of eccentric retrograde rotators}
\shortauthors{Kreyche, Barnes, Quarles, Lissauer, Chambers, $\&$ Hedman}
\usepackage{scalerel}
\usepackage{tikz}
\usetikzlibrary{svg.path}

\definecolor{orcidlogocol}{HTML}{A6CE39}
\tikzset{
 orcidlogo/.pic={
    \fill[orcidlogocol] svg{M256,128c0,70.7-57.3,128-128,128C57.3,256,0,198.7,0,128C0,57.3,57.3,0,128,0C198.7,0,256,57.3,256,128z};
    \fill[white] svg{M86.3,186.2H70.9V79.1h15.4v48.4V186.2z}
                svg{M108.9,79.1h41.6c39.6,0,57,28.3,57,53.6c0,27.5-21.5,53.6-56.8,53.6h-41.8V79.1z M124.3,172.4h24.5c34.9,0,42.9-26.5,42.9-39.7c0-21.5-13.7-39.7-43.7-39.7h-23.7V172.4z}
                svg{M88.7,56.8c0,5.5-4.5,10.1-10.1,10.1c-5.6,0-10.1-4.6-10.1-10.1c0-5.6,4.5-10.1,10.1-10.1C84.2,46.7,88.7,51.3,88.7,56.8z};
 }
}

\newcommand\orcidicon[1]{\href{https://orcid.org/#1}{\mbox{\scalerel*{
\begin{tikzpicture}[yscale=-1,transform shape]
\pic{orcidlogo};
\end{tikzpicture}
}{|}}}}

\usepackage[caption=false]{subfig}

\begin{document}

%\captionsetup[table]{labelsep=space}

\title{Retrograde-rotating exoplanets experience obliquity excitations in an eccentricity-enabled resonance}

\author{Steven M. Kreyche \orcidicon{0000-0002-7274-758X}}
\affiliation{Department of Physics, University of Idaho \\
Moscow, Idaho, USA}
\email{stevenkreyche@gmail.com}

\author{Jason W. Barnes \orcidicon{0000-0002-7755-3530}}
\affiliation{Department of Physics, University of Idaho \\
Moscow, Idaho, USA}

\author{Billy L. Quarles \orcidicon{0000-0002-9644-8330}}
\affiliation{Center for Relativistic Astrophysics, School of Physics, Georgia Institute of Technology  \\
Atlanta, Georgia, USA}

\author{Jack J. Lissauer \orcidicon{0000-0001-6513-1659}}
\affiliation{Space Science and Astrobiology Division, NASA Ames Research Center \\
Moffett Field, California, USA}

\author{John E. Chambers}
\affiliation{Department of Terrestrial Magnetism, Carnegie Institution of Washington\\
Washington, DC, USA}

\author{Matthew M. Hedman \orcidicon{0000-0002-8592-0812}}
\affiliation{Department of Physics, University of Idaho \\
Moscow, Idaho, USA}




\begin{abstract}
Previous studies have shown that planets that rotate retrograde (backwards with respect to their orbital motion) generally experience less severe obliquity variations than those that rotate prograde (the same direction as their orbital motion). Here we examine retrograde-rotating planets on eccentric orbits and find a previously unknown secular spin-orbit resonance that can drive significant obliquity variations. This resonance occurs when the frequency of the planet's rotation axis precession becomes commensurate with an orbital eigenfrequency of the planetary system. The planet's eccentricity enables a participating orbital frequency through an interaction in which the apsidal precession of the planet's orbit causes a cyclic nutation of the planet's orbital angular momentum vector. The resulting orbital frequency follows the relationship $f = 2 \dot{\varpi} - \dot{\Omega}$, where $\dot{\varpi}$ and $\dot{\Omega}$ are the rates of the planet's changing longitude of periapsis and ascending node, respectively. We test this mechanism by simulating cases of a simple Earth-Jupiter system, and confirm the predicted resonance. Over the course of 100 Myr, the test Earths with rotation axis precession rates near the predicted resonant frequency experienced pronounced obliquity variations of order $10^\circ$-$30^\circ$. These variations can be significant, and suggest that while retrograde rotation is a stabilizing influence most of the time, retrograde rotators can experience large obliquity variations if they are on eccentric orbits and enter this spin-orbit resonance.

\end{abstract}



\section{Introduction} \label{sec:intro}
The possibility of discovering extraterrestrial life largely motivates the study of exoplanets. Judging the habitability of these worlds requires the consideration of a multitude of factors. Aside from the basic requirement that a planet's orbit must reside within the habitable zone (HZ), the region around a star at which liquid water can exist on the planet's surface, the nature of the planet's obliquity, or axial tilt, is also significant to habitability. Planetary obliquity affects the climate of the planet by controlling the distribution and seasonal variation of its incoming solar flux. Therefore it is important to know the value of the planet's obliquity, and how it evolves over time. 

Formally, planetary obliquity, $\Psi$, is the angle between a planet's orbital and rotational angular momentum vectors. Therefore, with respect to the planet's direction of orbital motion, obliquity values $< 90^\circ$ correspond to prograde rotation while obliquity values $> 90^\circ$ correspond to retrograde (backwards) rotation. The Earth has a relatively low obliquity of $\sim 23.5^\circ$, in which its equator receives the most annually-averaged illumination while its poles receive very little. On the other hand, planets with obliquities near $90^\circ$ have poles that experience extreme contrasts in solar flux over the course of an orbit, while their equators actually receive the least amount of illumination as seen in Figure 1 of \citet{Lissauer_2012}. 

There are unique consequences for each value of obliquity. Low-obliquity worlds have the potential to enter snowball states, in which ice envelops the entire planet similar to the historic snowball Earth episode \citep{Hoffman_1998}. High obliquity worlds have a severe seasonality that can generally lead to warmer climates \citep{Kang_2019}. This seasonality could act to stave off snowball states \citep{Spiegl_2015, Colose_2019} or produce an equatorial ice belt \citep{Kilic_2018}. \citet{Olson_2019} found that oceans of high obliquity worlds may have more efficient nutrient recycling processes that would benefit potential biological activity near the surface. Ultimately the fate of a planet's climate depends on the configuration and properties of the planetary system, where studies that incorporate energy balance models show that any value of obliquity has the potential to provide habitable conditions \citep{Williams_1997, Williams_2003, Kilic_2017, Kane_2017, Guendelman_2019, Dong_2019, Colose_2019}.

Since the value of obliquity affects planetary climate, changes in the obliquity drive changes in the climate over time. In most cases, the expectation is that large swings in obliquity would likely be harmful to a planet's habitability, acting as jolts to its climate; for instance the large obliquity variations of Mars contributed to its atmospheric collapse \citep{Head_2004, Head_2005, Forget_2013}. The Earth experiences variations in its obliquity of just $\sim 2.4^\circ$, which drives glacial cycles and Ice Ages \citep{Milankovic_1998}.  However, \citet{Armstrong_2014} found that in special circumstances extreme obliquity variations can be helpful, in that they can push the outer limit of the HZ outwards by staving off snowball states. Later, \citet{Deitrick_2018b} applied a more robust model and argued that fast and large variations in a planet's rotational and orbital properties can actually do the opposite, and lead to global glaciation.

The value of a planet's orbital eccentricity, $e$, also influences the nature of its climate. Conservation of angular momentum requires an eccentric planet to spend more time near its apoapsis (the orbit's furthest point from the primary body) than its periapsis (the closest point to the primary body). At first glance, this relationship seems to imply that significantly eccentric planets should be inhospitable, where a greater amount of time spent further away from their energy source would trigger global glaciation. However, the time averaged solar flux over the course of an orbit actually works out to 

\begin{equation}
\langle F \rangle \propto (1-e^2)^{-\frac{1}{2}}
\end{equation}

\noindent This relationship shows that a planet actually receives a larger orbitally-averaged global flux with increasing values of eccentricity \citep{Laskar_1993a}, albeit the planet's average equilibrium temperature decreases slightly \citep{Mendez_2017}. Multiple studies applied energy balance and global circulation models to investigate the viability of eccentric planets as habitable worlds. \citet{Williams_2002} found that planets up to $\sim 0.7$ eccentricity could remain habitable even with seasonal departures outside of the HZ. Later studies generally confirmed these findings with the exception of the occurrence of snowball states for the case of planets orbiting stars hotter than the Sun \citep{Dressing_2010, Bolmont_2016}. The climate of an Earth-like world could even remain temperate throughout significant eccentricity variations over short timescales, as demonstrated by  \citet{Way_2017}. Therefore, we should not necessarily discard eccentric worlds as potentially habitable candidates and instead should study the effects of a planet's orbital eccentricity in conjunction with its obliquity.

Terrestrial planet obliquities are likely initially isotropic in distribution, randomly decided from collisions in the protoplanetary disk after formation \citep{Dones_1993, Lissauer_1997, Miguel_2010}, see however \citet{Lissauer_1991}. However, \citet{Millholland_2019} found that planet-disk interactions can act to influence planetary obliquity early on. Venus and Uranus in our own Solar System are retrograde rotators. Based on these observations, obliquity studies should consider the entire range of possible obliquity values. 

A handful of studies considered the obliquity evolution of retrograde rotators. \citet{Laskar_1993b} studied the obliquity evolution of the Earth under the influence of the Moon, and reasoned that cases of Earths with retrograde obliquities would be expected to be more stable than prograde ones. Later, \citet{Lissauer_2012} explored the obliquity variations of a moonless Earth and found in agreement, that the retrograde-rotating Earths were generally more obliquity stable. \citet{Barnes_2016} then explored the case of an early Venus and again reported similar results, with the exception of a long-term pronounced variability for some retrograde rotators. \citet{Quarles_2019} found that depending on the mutual inclination and orbital precession of the bodies, retrograde rotators in the $\alpha$ Centauri AB binary-star system would likely be especially obliquity stable. Together, these studies have shown that retrograde rotation largely stabilizes obliquity under most circumstances. 

In this paper we present an exceptional circumstance in which a spin-orbit resonance enabled by a retrograde-rotating planet's orbital eccentricity can drive it to experience significant obliquity variations. Specifically, the planet's eccentricity triggers a complex mechanism that enables a 1:1 secular spin-orbit resonance, in which the frequency of the planet's rotation axis precession becomes commensurate with an orbital eigenfrequency of the system. 

In this work, we explore the obliquity stability of retrograde rotators, finding that orbital eccentricity can generate large obliquity variations. We begin with Section \ref{sec:conceptual}, in which we describe the mechanism that affects the obliquity of retrograde rotators. Then in Section \ref{sec:numerical} we explain our approach to test this mechanism. In Section \ref{sec:results}, we reveal the results of our frequency analysis and simulations, and discuss their implications. We summarize our results and their implications for habitability in Section \ref{sec:conclusion}.

\section{Conceptual Model} \label{sec:conceptual}
Planetary obliquity, $\Psi$, is the angle between a planet's orbital and rotational angular momentum vectors. Therefore, a change in the orientation of \emph{either} of these vectors will alter a planet's obliquity. Torques exerted on a planet's rotational bulge from its star and neighboring planets can directly change the orientation of the planet's rotational angular momentum vector. On the other hand, changes in a planet's orbital inclination, $I$, or changes in the its longitude of ascending node, $\Omega$, the position along an inclined orbit at which a planet ascends from below the reference plane to above \citep{Murray_2000}, alter the orientation of the planet's orbital angular momentum vector. Aside from the most basic case of a single-planet system, the properties and orientation of a planet's orbit in a multi-planet system evolve over time due to gravitational interactions with neighboring bodies. 

Within this section, we first discuss how a planet's rotation axis evolves over time in Section \ref{sec:rotevo}. Then we review the different ways that a planet's orbit can evolve in Section \ref{sec:orbitevo}. Finally, Section \ref{sec:eccmech} describes how these processes can act in conjunction to lead to excitations of the obliquities of retrograde rotators.

\subsection{Rotation Axis Evolution} \label{sec:rotevo}
Two angles fully describe the orientation of a planet's rotational angular momentum vector: the obliquity and the precession angle, $\mathbf{\phi}$. While the obliquity specifies the polar angle between the orbital and rotational angular momentum vectors, the precession angle specifies the azimuthal angle at which the rotational angular momentum lies about the orbital angular momentum vector. Torques exerted on the planet's equatorial bulge by the primary body and/or by potential satellites cause this angle to precess over time. This phenomenon is rotation axis precession, or axial precession. Axial precession acts in the opposing direction to the orbital motion for planets with prograde obliquities. For planets with retrograde obliquities however, the direction of precession flips, as shown in Figure \ref{fig:nodal_apsidal}, and instead moves in the direction of orbital motion. Following \citet{Surgy_1997}, the planetary obliquity, precession constant ($\alpha$), and orbital eccentricity define the rate of the axial precession as

\begin{equation}
    \dot{\phi} = \frac{\alpha \cos{\Psi} }{(1-e^2)^{\frac{3}{2}}}
    \label{eqn:prec_freq}
\end{equation}

\noindent The precession constant is a function of the planet's mean motion, $n$, the zonal harmonic constant related to planetary oblateness, $J_2$, the planet's rotational frequency, $\mathbf{\nu}$, and a constant related to the planet's moment of inertia, $\overline{C}$. This takes the form

\begin{equation}
    \alpha \approx \frac{3n^2}{2\nu} \frac{J_2}{\overline{C}}
    \label{eqn:prec_const}
\end{equation}

\noindent This relationship is only approximate due to the assumption that a planet's dynamical ellipticity is roughly proportional to $\nu^2$ \citep{Laskar_1993b, Li_2014, Quarles_2020}.

\subsection{Orbital Evolution} \label{sec:orbitevo}
\begin{figure}[ht!]
\centering
\includegraphics[width=\textwidth]{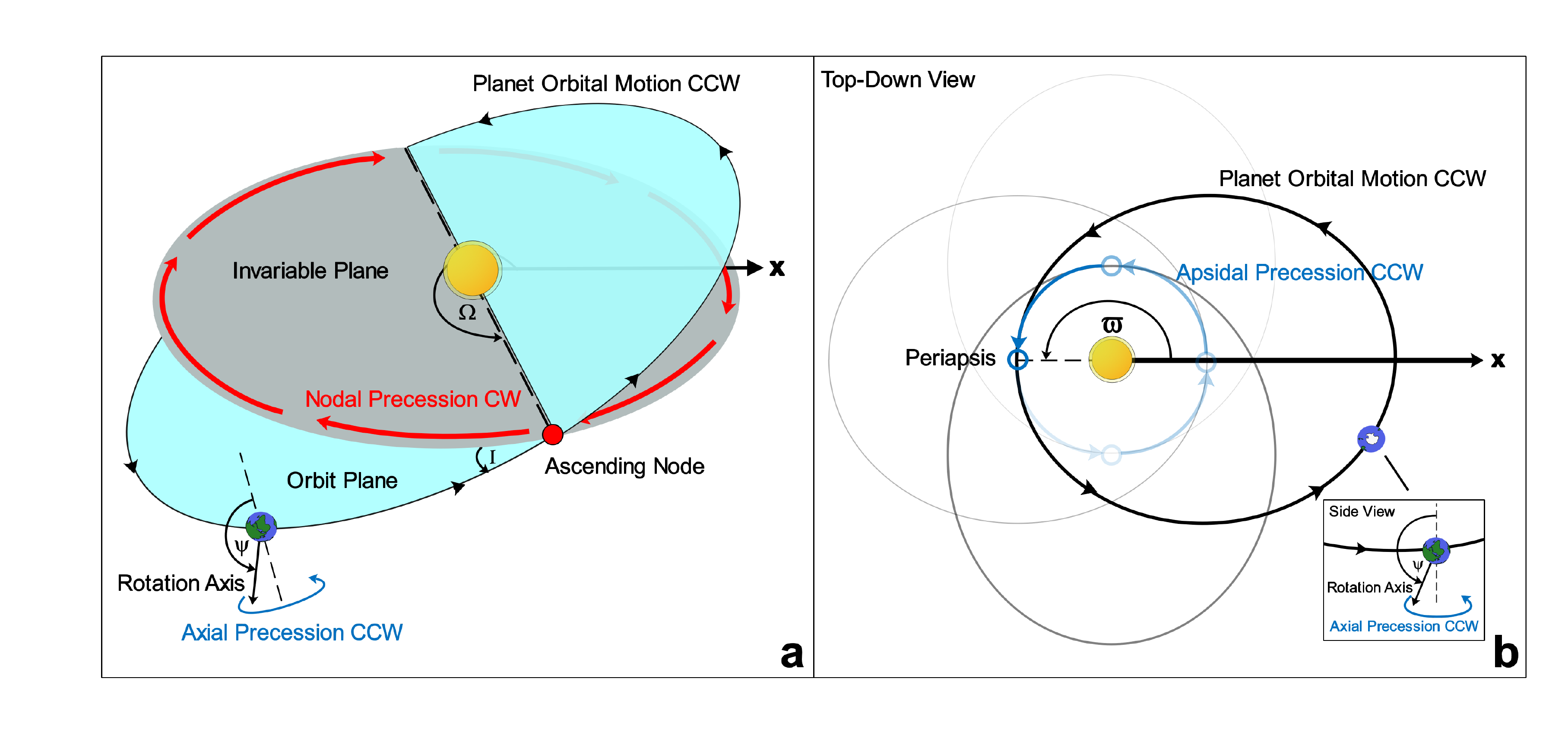}
\caption{Panel (a) depicts a visualization of the nodal precession of a retrograde-rotating planet's orbit that has some inclination, $I$, with respect to the invariable plane (the plane that lies orthogonal to the net angular momentum vector of the planetary system). During this process, the longitude of the planet's ascending node, $\Omega$, precesses in the clockwise direction while the planet's rotation axis precesses in the opposite (counter-clockwise) direction. Panel (b) depicts a visualization of the apsidal precession of the eccentric planet's orbit with a top-down view situated above the orbit plane. During this process the planet's longitude of periapsis, $\varpi$, precesses in the counter-clockwise (prograde) direction while the planet's rotation axis precesses in the same direction, as seen with the side view within the figure inset.}
\label{fig:nodal_apsidal}
\end{figure}

\begin{figure}[!tbp]
  \centering
  \subfloat[Nodal Precession]{\includegraphics[width=0.475\textwidth]{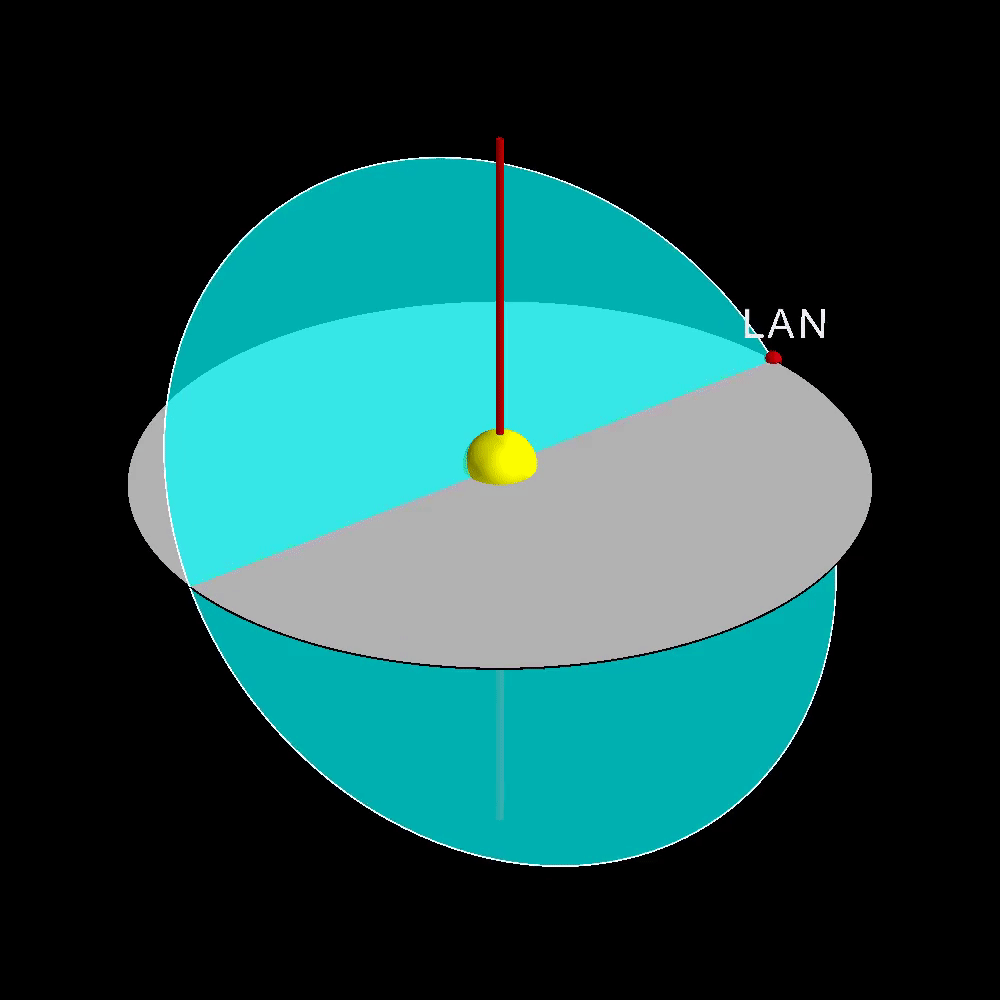}\label{fig:ani_nodal}}
  \hfill
  \subfloat[Apsidal Precession]{\includegraphics[width=0.475\textwidth]{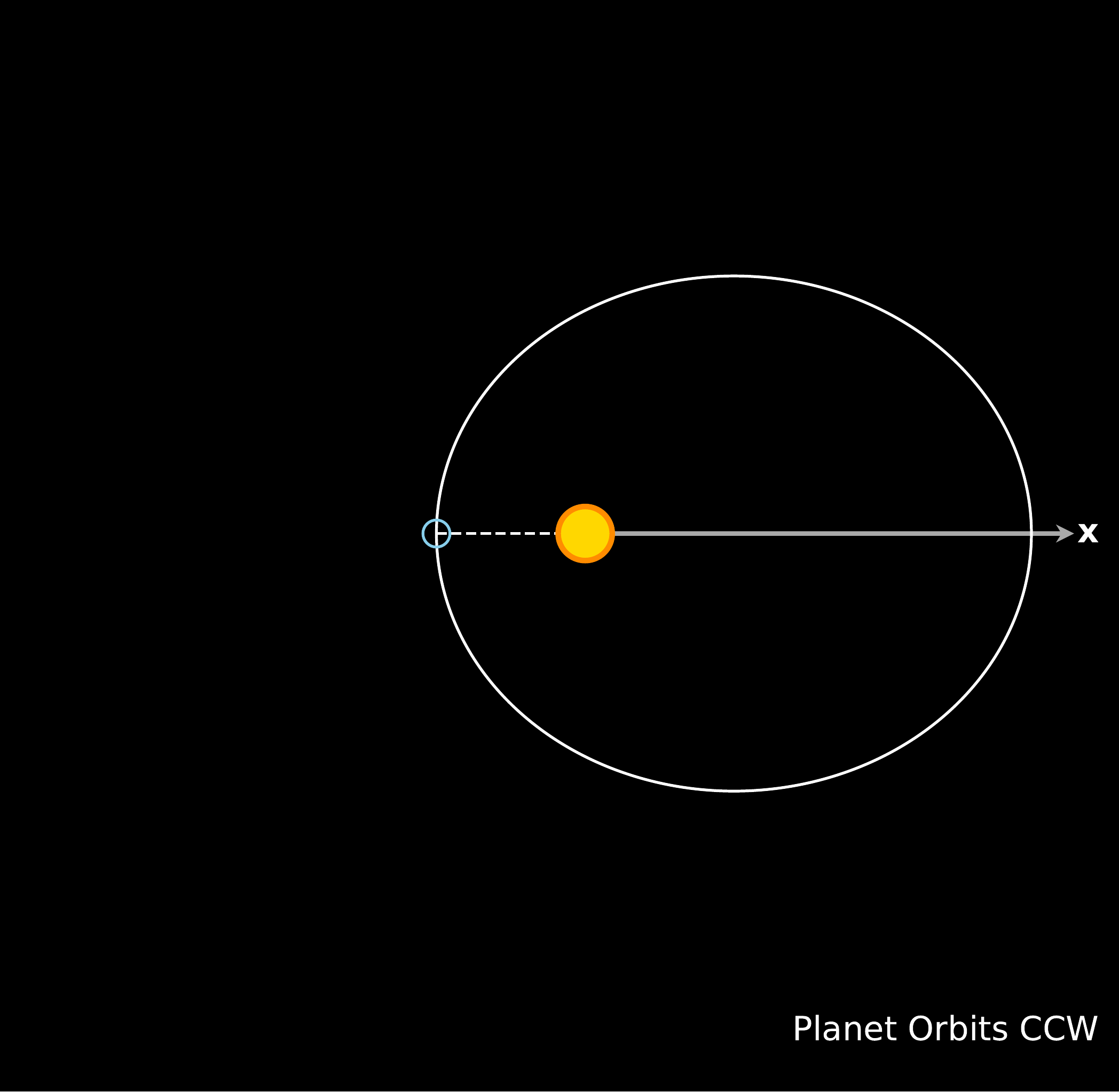}\label{fig:ani_apsidal}}
  \caption{An animation of this figure is available at https://github.com/SMKreyche/Kreyche-et-al.-2020-Videos. The real-time duration of the video is 14 seconds, in which similar to to panel (a) of Figure \ref{fig:nodal_apsidal}, Figure \ref{fig:ani_nodal} depicts one cycle of the nodal precession of a planet's cyan-colored orbit which is inclined with respect to the grey invariable plane (the plane that lies orthogonal to the net angular momentum of the planetary system). During this process we show the planet's ascending node as a red ball located at the orbital longitude of the ascending node (LAN or $\Omega$ within this text), which precesses in the clockwise direction. Figure \ref{fig:ani_apsidal} is similar to panel (b) of Figure \ref{fig:nodal_apsidal}, and depicts one cycle of a top-down view of the apsidal precession of an eccentric planet's orbit. During this process, the planet's periapsis (the point along its orbit that is closest to its star shown as an open blue circle) precesses in the counter-clockwise direction.}
\end{figure}

Considering the case of a multi-planet system, one way that a planet's orbit evolves over time is by nodal precession. Depicted in both panel (a) of Figure \ref{fig:nodal_apsidal} and Figure \ref{fig:ani_nodal}, nodal precession is the process in which the orbital ascending node precesses through space in the opposing direction to that of the planet's orbital motion. This corresponds to a cyclically changing direction at which the orbital angular momentum vector points in space with respect to the invariable plane (the plane that lies orthogonal to the net angular momentum vector of the planetary system) at the frequency $\dot{\Omega}$ (the rate of the changing longitude of ascending node). This process resembles a coin spinning on a table top, in which the top face of the coin reorients itself in space as it rotates. Physically, forces from neighboring planets that are normal to a planet's orbit plane as it traverses along its orbit drive nodal precession, with a complete cycling occurring on a timescale of $\sim 70,000$ years for the Earth \citep{Muller_1995}. Therefore this process only occurs when either the orbital plane of the planet shares a mutual inclination with that of a neighboring planet, or their orbital ascending nodes share some angular separation.

% Figure that shows nodal precession and apsidal precession of retrograde-rotating Earth

Looking to both panel (b) of Figure \ref{fig:nodal_apsidal} and Figure \ref{fig:ani_apsidal}, for the case of an eccentric orbit, the orientation of a planet's orbit can also evolve through a process called apsidal precession. This process involves the precession of the line of apsides, the imaginary line connecting the orbit's periapsis and apoapsis. During apsidal precession the orbital argument of periapsis, $\omega$, the angle from the orbital ascending node to the position of periapsis, as well as the orbital longitude of periapsis, $\varpi$, the angle from the origin of the reference frame to the position of periapsis, both precess in the same direction as that of the planet's orbital motion.  The motion of apsidal precession ``hula hoops" the planet's orbit around its star at the frequency $\dot{\omega}$ (the rate of the changing argument of periapsis). In contrast to nodal precession, forces from neighboring planets that are radial and tangential to the planet along its orbit drive apsidal precession. This precession cycle elapses over the course of $\sim 112,000$ years for the Earth \citep{Heuvel_1965}. Unlike nodal precession, apsidal precession does not affect obliquity directly because the orientation of the orbital angular momentum vector does not change.

\subsection{A Spin-Orbit Resonance Enabled by Eccentricity} \label{sec:eccmech}
Bearing in mind the processes in which the rotation axis and planetary orbit can evolve, special circumstances can induce large-amplitude obliquity variations. One such circumstance is the event of a 1:1 secular spin-orbit resonance between a planet's axial precession frequency, $\dot{\phi}$, and an orbital eigenfrequency of the planetary system (usually a driver of nodal precession). While prograde rotators experience axial precession in the same direction as their nodal precession and can potentially achieve this spin-orbit resonance, retrograde rotators conventionally cannot. A positive orbital frequency that comes near a retrograde rotators axial precession frequency does not usually exist. Indeed, previous studies found retrograde rotators to be especially obliquity stable \citep{Laskar_1993b, Lissauer_2012, Barnes_2016}. We propose an exceptional circumstance. The orbital eccentricity of a retrograde-rotating planet can act to trigger a mechanism that enables a spin-orbit resonance which consequently drives obliquity variations. 

\begin{figure}[ht!]
\centering
\includegraphics[width=\textwidth]{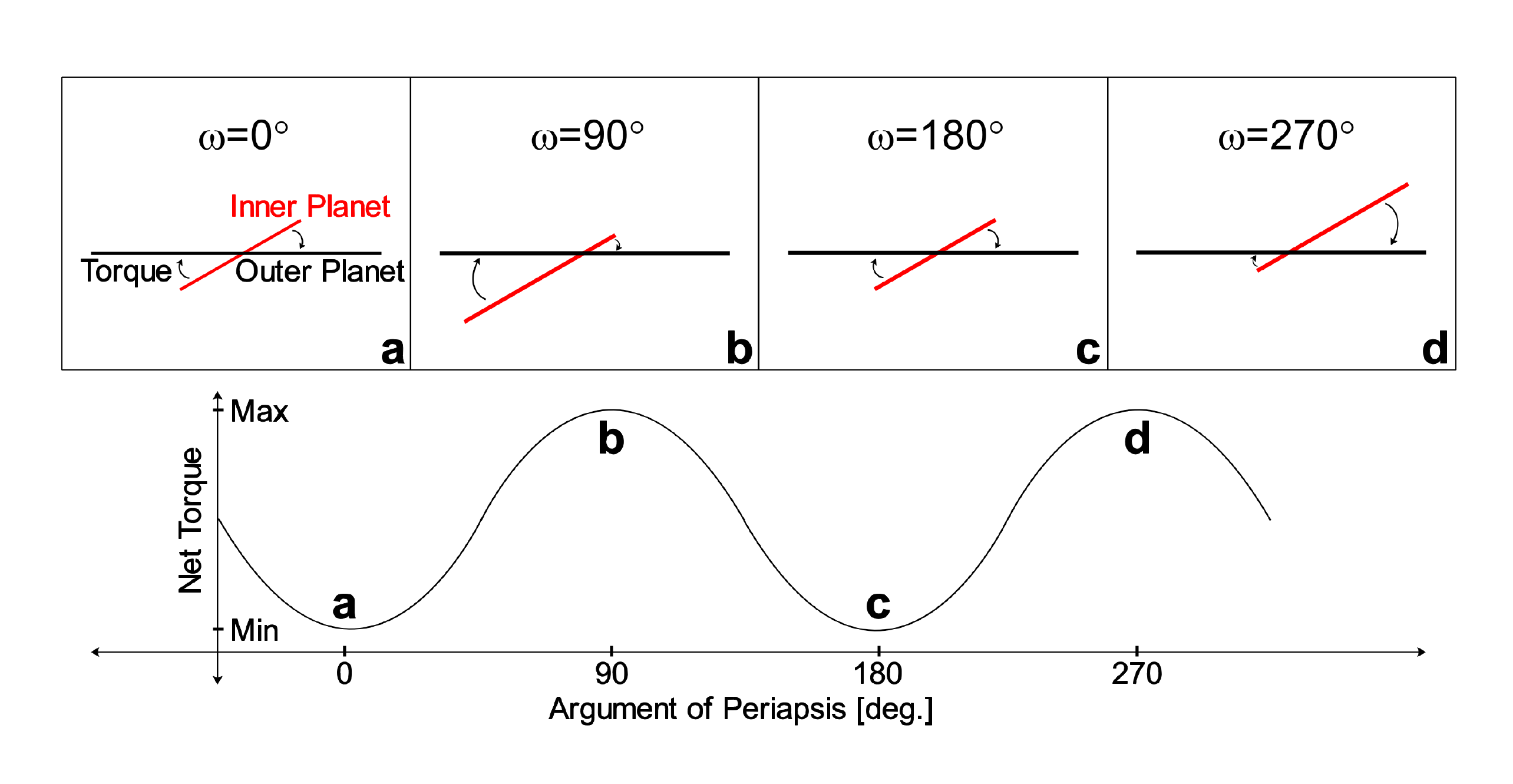}
\caption{This edge-on perspective shows a massive outer planet's orbital plane (black) and an inclined, eccentric inner planet's orbital plane (red). The panels show four cases of the inner planet's argument of periapsis, $\omega$, as it undergoes apsidal precession. Here we ignore the nodal precession of their orbits for simplicity and assume their longitudes of ascending node are equal (their orbital planes both remain edge on from our perspective). The cases for $\omega = 0^\circ$ and $180^\circ$ (when the line of apsides is co-planar with outer planet's orbital plane) occur twice, for which the normal force felt by the inner planet is zero at periapsis and apoapsis. The cases for $\omega = 90^\circ$ and $270^\circ$ show that the inner planet's orbital plane is tilted out of the outer planet's orbital plane, enabling a precessional torque due to the stark contrast in the normal force felt by the inner planet at the apsides. The letters on each of the panels map to the plot beneath, showing the net torque on the inner planet's orbit as a function of $\omega$. The difference between the minimum and maximum net torque would grow for increasing eccentricity.}
\label{fig:mechanism}
\end{figure}

For a planet with nonzero eccentricity, the process of apsidal precession indirectly affects that of nodal precession and introduces an additional orbital frequency. Figure \ref{fig:mechanism} demonstrates this mechanism and shows four edge-on snapshots of a two-planet system, each with a different value of the inner planet's argument of periapsis. Here the inner planet has some mutual inclination with the outer body, and its orbit experiences apsidal precession as $\omega$ changes. Considering a more massive exterior planet which can be thought of as a uniform ring of mass in the long-timescale secular approximation, the orbit of the inner planet tips asymmetrically out of the ring mass plane twice over the duration of the precession cycle, that is, at a frequency of $2\dot{\omega}$. The inner planet therefore feels a difference in force pulling on it and its orbit experiences a precessional torque as it proceeds over the course of this cycle. This is important not only due to the contrast between the distance of the orbit's periapsis and apoapsis, but also because conservation of angular momentum requires that more time is spent at apoapsis compared to periapsis. This effect becomes more and more significant with larger values of eccentricity, and affects the process of nodal precession in the sense that the orbital angular momentum vector experiences an additional nutating motion. 

\begin{figure}[!tbp]
  \centering
  \subfloat[The Physical Process]{\includegraphics[width=0.57\textwidth]{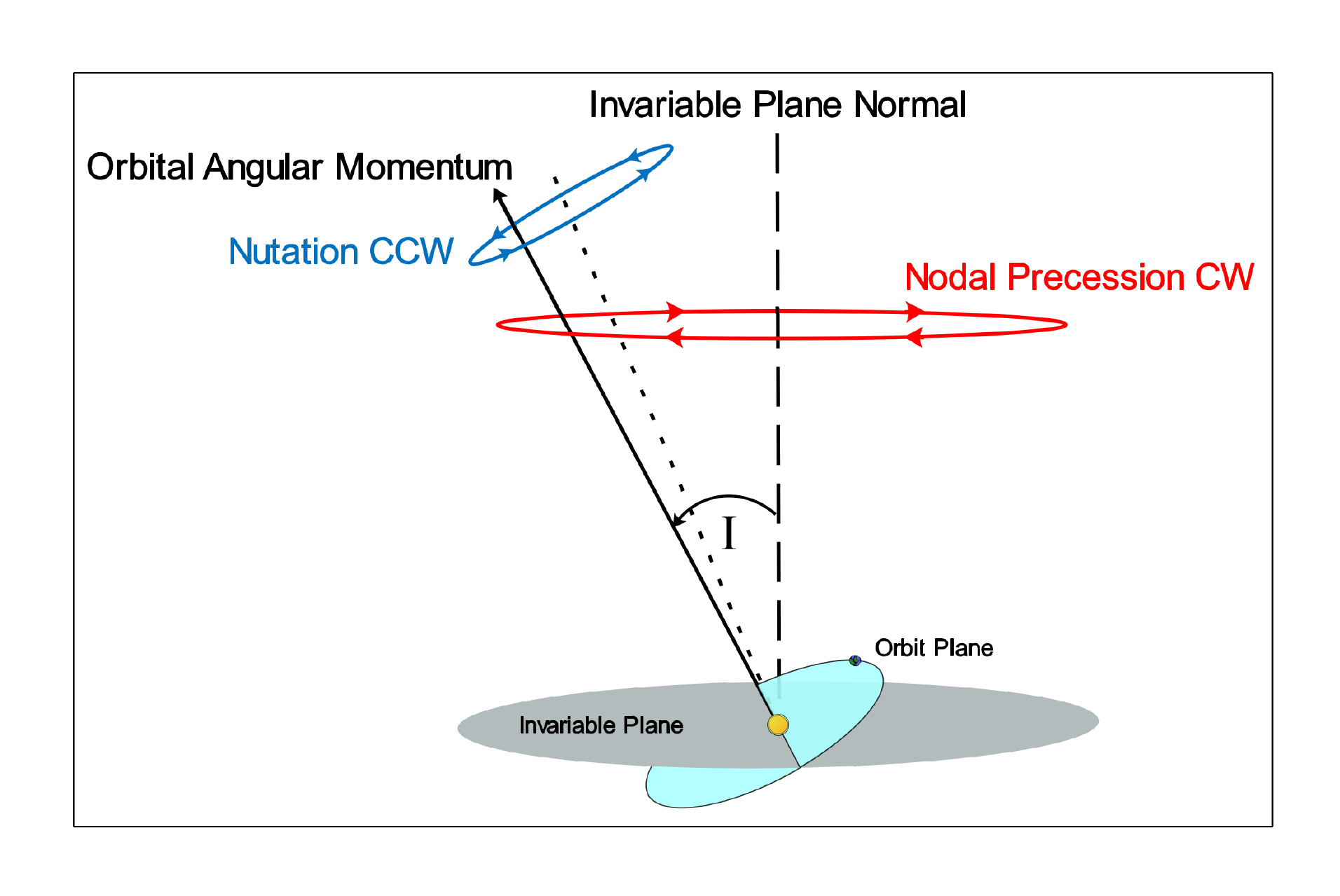}\label{fig:nutation}}
  \hfill
  \subfloat[Superimposed Precessions]{\includegraphics[width=0.38\textwidth]{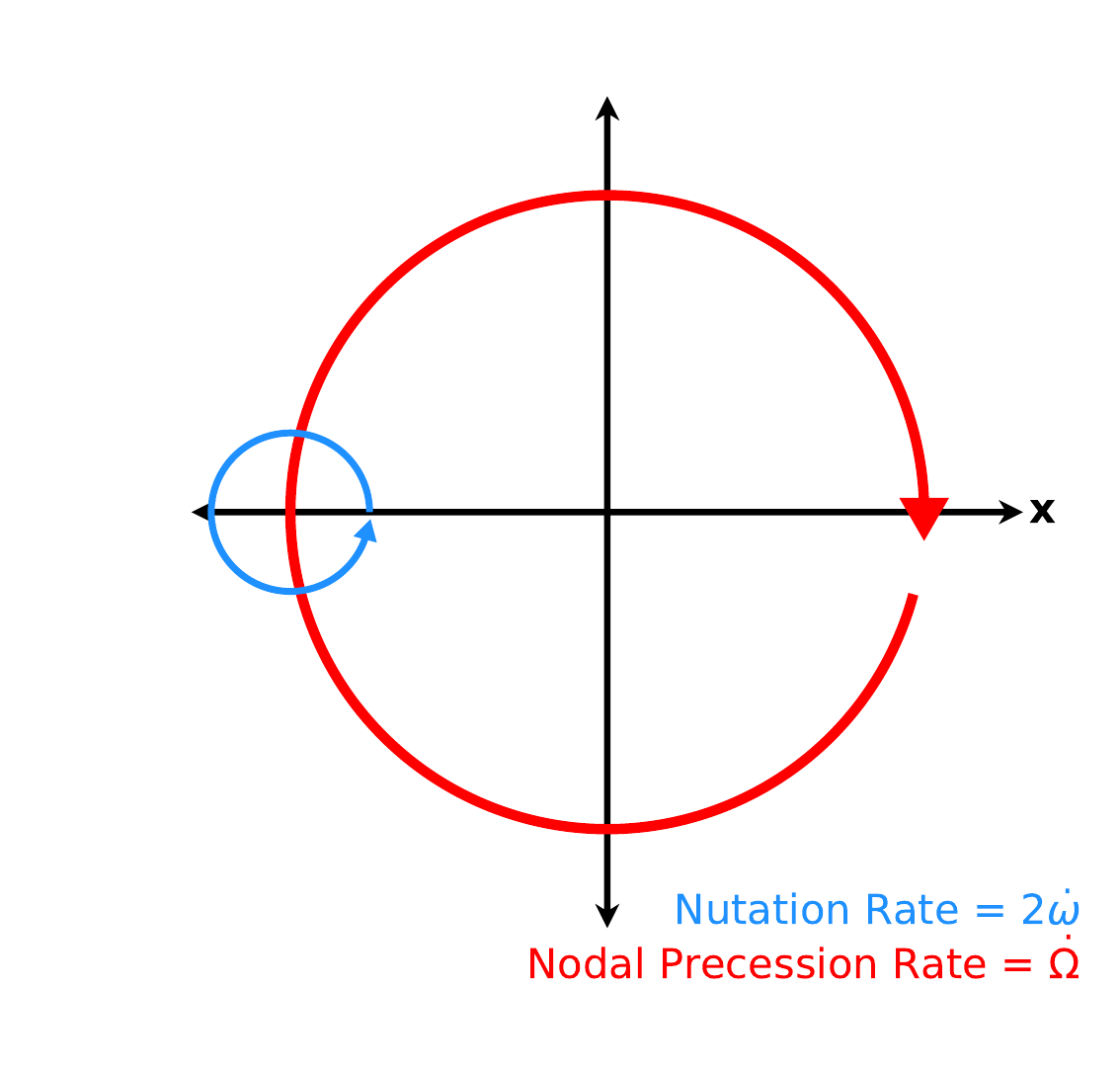}\label{fig:precessions}}
  \caption{Figure \ref{fig:nutation} shows a planet's orbital angular momentum vector as the straight black vector arrow. The planet's orbit has some inclination, $I$, with respect to the invariable plane (the plane that lies orthogonal to the net angular momentum of the planetary system). Due to the perturbations of a neighboring planet, its orbital angular momentum vector precesses around the invariable plane normal in the clockwise direction (viewed from above). In the case that the planet's orbit is eccentric, its orbital angular momentum vector nutates in the counter-clockwise direction simultaneously. An animation of Figure \ref{fig:precessions} is available at https://github.com/SMKreyche/Kreyche-et-al.-2020-Videos. The real-time duration of the video is 11 seconds, in which we show one cycle of the two processes that we describe in Figure \ref{fig:nutation} acting as superimposed precessions with rates of $\dot{\Omega}$ and $2\dot{\omega}$, where $\dot{\Omega}$ and $\dot{\omega}$ are the rates at which the longitude of ascending node and argument or periapsis are changing, respectively.}
\end{figure}

Figure \ref{fig:nutation} visualizes this process, where an opposing nutating circular motion accompanies the traditional circular motion traced out by the orbital angular momentum vector during nodal precession. This motion can be thought of as two superimposed precessions, each with its own amplitude, frequency, and direction. Figure \ref{fig:precessions} summarizes this concept and shows that the primary nodal precession frequency drives the orbital angular momentum vector at the frequency $\dot{\Omega}$ while the nutating motion operates at the frequency $2\dot{\omega}$. The resulting orbital frequency we expect to see for this nutation goes as

\begin{equation}
f = 2 \dot{\omega} + \dot{\Omega} = 2 \dot{\varpi} - \dot{\Omega}
\label{eqn:predicted_freq}
\end{equation}

\noindent where we used the relationship $\varpi = \Omega + \omega$ which translates to $\dot{\varpi} = \dot{\Omega} + \dot{\omega}$. Following this, the condition for the aforementioned spin-orbit resonance is when $f = \dot{\phi}$, or $2 \dot{\varpi} - \dot{\Omega} - \dot{\phi} = 0$.

% Figure that shows edge-on view of Earth-Jupiter system for 4 cases of omega

\section{Numerical Treatment} \label{sec:numerical}

\subsection{Approach}
We explore the validity of our analytical derivation by performing numerical simulations with the use of the mixed-variable symplectic N-body integrator, \texttt{SMERCURY}, which is a modified algorithm that adds spin-tracking capabilities to the original \texttt{MERCURY} package \citep{Chambers_1999}. Similar to the works of \citet{Lissauer_2012}, \citet{Barnes_2016}, and \citet{Quarles_2020}, we use \texttt{SMERCURY} to track a planet's obliquity evolution while the planetary system dynamically evolves. We refer to \citet{Lissauer_2012} for a thorough description of our technique, although we highlight some key details here.

\texttt{SMERCURY} computes the orbital evolution of the system's bodies based on orbital forcing interactions between one another, in which we base the calculations on a time step of at most $5\%$ of the orbital period of the inner-most planet. In the meantime, the $\texttt{SMERCURY}$ algorithm only tracks the obliquity evolution of one body in a specified system, although it features what we call ``ghost planets". Ghost planets are essentially massless clones of the tracked body with their own assigned rotation states, in which their obliquities evolve independently. This scheme allows for a broad exploration of parameter space and saves computation time. 

A $\texttt{SMERCURY}$ simulation treats the tracked planet along with its ghost planets as axisymmetric rigid bodies, in which their rotation states evolve due to gravitational torques exerted on their rotational bulges. We assign a value for the planet's oblateness coefficient $J_2$ based on the their rotation periods, computed by following the Darwin-Radau relation \citep{Hubbard_1984, Murray_2000}. This value remains fixed throughout the duration of the simulation, which implies that we do not account for tidal dissipation effects. We can neglect these effects so long as the timescale for tidal effects to become important for the planet greatly exceeds the chosen integration time. Considering a moonless Earth with a rotation period of 24 hours as an example, it would take of order $10^{10}$ years for the Earth to become tidally locked with the Sun. In this case, the amount of tidal deceleration over the course of a 100 Myr simulation would have little effect on its spin evolution.

We initialize each simulation by generating the rotation states of each planet based on their desired orientation and rotation period, taking into account the orientation of the orbit plane. Recalling from Section \ref{sec:conceptual}, together the obliquity and the precession angle specify the orientation of a planet's rotational angular momentum vector. We transform these angles by accounting for the orbital inclination and longitude of ascending node, following the explanation of \citet{Surgy_1997} and \citet{Barnes_2016}.

\subsection{Initial Conditions} \label{sec:initial}
We design a simple numerical experiment to test the eccentricity mechanism that we predict destabilizes the obliquities of retrograde rotating planets. We consider a toy system consisting of the Sun as the primary body, a moonless Earth we call Earthmoo (the exclusion of the Moon simplifies our experiment), and Jupiter. Earthmoo is taken to have the same density as the real Earth yet has the mass of the combined Earth-Moon system, which is consistent with the approach of \citet{Lissauer_2012} and \citet{Barnes_2016}. We arbitrarily initialize our simulations according to Table \ref{tab:orbital}, for which the orbital elements are with respect to the J2000 epoch and ecliptic \citep{Murray_2000}. Earthmoo's orbital elements mostly mimic that of the real Earth-Moon barycenter at this epoch, although we assign values of 0, 0.1, 0.3, and 0.5 for its initial orbital eccentricity across separate simulations. We begin Jupiter on a circular orbit (its initial eccentricity is set to zero) at the start of all of our simulations in order to avoid unwanted excitation of Earthmoo's eccentricity. Importantly, we place a mutual inclination between Earthmoo and Jupiter in order to ensure that the orbit of Earthmoo will experience nodal precession as Jupiter exerts gravitational tugs normal to Earthmoo's motion. We set the initial orbital inclinations of Earthmoo and Jupiter to the arbitrary values of $2.30530^\circ$ and $1.30530^\circ$, respectively; note that their mutual inclination actually works out to be $3.04204^\circ$ from the spherical law of cosines, due to the angular difference between their orbital ascending nodes. \citet{Deitrick_2018a} similarly tested a mutually inclined Earth-Jovian system with large eccentricities, but did not test Earths with retrograde obliquities.

% table that includes EJ system orbital parameters
\begin{deluxetable}{c c c c c c c c }[ht!]
\tabletypesize{\footnotesize}
\tablecaption{\\ System Orbital Parameters \label{tab:orbital}}
\tablehead{\colhead{Planet} & \colhead{$m$ [$M_{\odot}$]} & \colhead{$a$ [AU]} & \colhead{$e$} & \colhead{$I$ [deg.]} & \colhead{$\omega$ [deg.]} & \colhead{$\Omega$ [deg.]} & \colhead{$M$ [deg.]}}
\startdata
Earthmoo & 3.04 x $10^{-6}$ & 1.00000011 & - \tablenotemark{b} & 2.30530 & 114.20783 & 348.73936 & 357.51716 \\
Jupiter & 9.5450 x $10^{-4}$ & 5.20336301 & 0 & 1.30530 & 274.19770 & 100.55615 & 19.65053 \\
\enddata
\tablecomments{Initial values for the plantary mass ($m$), semimajor axis ($a$), eccentricity ($e$), inclination ($I$), argument of periapsis ($\omega$), longitude of ascending node ($\Omega$), and mean anomomly ($M$) for the Earthmoo-Jupiter system taken from \citet{Murray_2000}. These values correspond to the J2000 ecliptic with respect to the Earth-Moon barycentre. Mass of Sun taken to be 1.98911$\times 10^{30}$ kg.}
\tablenotetext{b}{We set Earthmoo's eccentricity to initial values of 0, 0.1, 0.3, and 0.5.}
\end{deluxetable} 

We explore a range of rotation states composed of varied rotation periods and initial obliquities that correspond to varied axial precession frequencies following Equation \ref{eqn:prec_freq}. This variation allows for a contrast of Earthmoos caught both in and out of our proposed spin-orbit resonance. Therefore, we test a range of Earthmoo rotation periods spanning 4 to 46 hours, where we select the 4 hour lower limit due to the physical limitation that Earthmoo would be near breakup and no longer axisymmetric, sticking to our Darwin-Radau assumptions discussed in the previous section. Taking Earthmoo to have a density of 5.5153 g/cm$^3$ and a constant moment of inertia coefficient of 0.3296108, we compute the $J_2$ values for each case. We show a summary of the rotational parameters we use in Table \ref{tab:rotational}. Since this work investigates retrograde rotating planets, we test initial obliquities of Earthmoo ranging from $90^\circ$ to $180^\circ$, in $5^\circ$ increments. Each obliquity value that we test pairs with a precession angle ($\phi$) value; we arbitrarily set the initial $\phi$ value of the Earthmoos to $348.74^\circ$.

% table that includes EJ system rotational parameters
\begin{deluxetable}{c c c c}[ht!]
\tabletypesize{\footnotesize}
\tablecaption{\\ System Rotational Parameters \label{tab:rotational}}
\tablehead{\colhead{$P_{rot}$ [hr]} & \colhead{$R_{eq}$ [km]} & \colhead{$J_2$} & \colhead{$\alpha$ [$''$/yr]}}
\startdata
4 & 6721.04 & 0.0441685 & 118.86406 \\
6 & 6520.71 & 0.0179269 & 72.36607 \\
10 & 6438.62 & 0.0062130 & 41.80041 \\
16 & 6412.76 & 0.0023978 & 25.81175 \\
24 & 6403.80 & 0.0010612 & 17.13579 \\
34 & 6400.24 & 0.0005279 & 12.07568 \\
46 & 6398.64 & 0.0002882 & 8.91882 \\
\enddata
\tablecomments{Computed initial values for the rotation period ($P_{rot}$), equatorial radius ($R_{eq}$), oblateness coefficient value ($J_2$), and precession constant ($\alpha$) for Earthmoo determined from \citet{Lissauer_2012}.}
\end{deluxetable}

\section{Results and Discussion} \label{sec:results}

\subsection{Frequency Analysis} \label{sec:FMFT}
The mechanism that we describe in Section \ref{sec:conceptual} suggests that our Earthmoo-Jupiter system will involve a significant positive orbital frequency pertaining to the nodal precession of an eccentric Earthmoo's orbit. We predict that this frequency will take on a value according to Equation \ref{eqn:predicted_freq}. In order to test this mechanism, we perform a set of simulations and conduct a Fourier analysis to obtain the secular eigenfrequencies of the system. 

We ran 100 Myr simulations sampled at 100 year intervals of the Earthmoo-Jupiter system described by Table \ref{tab:orbital}; here we set the integration time step to $2.5\%$ of Earthmoo's orbital period. From this, we performed a Fourier analysis over the course of the simulation of the inclination vector ([$I\cos{\Omega}$, $I\sin{\Omega}$]) and the eccentricity vector ([$e\cos{\varpi}$, $e\sin{\varpi}$]). This is done using a Frequency Modified Fourier Transform \citep{Sidlichovsky_1996}, yielding the amplitude, frequency, and phase for each mode. The two most prominent amplitude frequencies (the subsequent frequencies have orders of magnitude less power and are thus left out) are displayed for each in Table \ref{tab:freqs}. We exclude a prominent peak of inclination vector found in each analysis at 0 $''$/yr, as it is an artifact due to a degeneracy in the inclination vector \citep{Murray_2000}.

% Table for the inclination vector orbital frequencies
\begin{deluxetable}{c | c | ccc | ccc }[ht!]
\tabletypesize{\footnotesize}
\tablecaption{\\ Secular Orbital Frequencies \label{tab:freqs}}
\tablehead{\colhead{} & \colhead{} & \colhead{} & \colhead{Inclination Vector} & \colhead{} & \colhead{} & \colhead{Eccentricity Vector} & \colhead{} \\
\hline
\colhead{$e_0$} & \colhead{$i$} & \colhead{$f_i$ [$''$/yr]} & \colhead{$A_i$} & \colhead{$\gamma_i$ [deg.]} & \colhead{$g_i$ [$''$/yr]} & \colhead{$B_i$} & \colhead{$\beta_i$ [deg.]}  }
\startdata
0 & 1 & -7.06 & 3.038000 & 325.28 & 7.03 & 0.000027 & 90.91 \\ 
0 & 2 & -14.13 & 0.000610 & 100.43 & 0.01 & 0.000003 & 247.14 \\ 
\hline
0.1 & 1 & -7.22 & 3.036219 & 325.7 & 7.00 & 0.100044 & 103.08\\
0.1 & 2 & 21.22 & 0.020543 & 240.49 & -21.45 & 0.000188 & 8.25 \\
\hline
0.3 & 1 & -8.49 & 3.036676 & 328.68 & 6.75 & 0.300010 & 103.07 \\
0.3 & 2 & 21.98 & 0.180591 & 237.46 & -23.72 & 0.000501 & 14.30 \\
\hline
0.5 & 1 & -10.95 & 3.110164 & 334.51 & 6.18 & 0.499828 & 103.05 \\
0.5 & 2 & 23.32 & 0.498464 & 231.59 & -28.08 & 0.000698 & 25.97 \\
\enddata
\tablecomments{The results from our Frequency Modified Fourier Transform analysis of the inclination vector ([$I\cos{\Omega}$, $I\sin{\Omega}$]) and the eccentricity vector ([$e\cos{\varpi}$, $e\sin{\varpi}$]) of Earthmoo for different initial values of eccentricity ($e_0$), where $I$ is the orbital inclination, $\Omega$ is the longitude of ascending node, $e$ is the eccentricity, and $\varpi$ is the longitude of periapsis. We show the top two values of the frequency ($f_i$), amplitude ($A_i$), and phase ($\gamma_i$) of the inclination vector, as well as the top two values of the frequency ($g_i$), amplitude ($B_i$), and phase ($\beta_i$) of the eccentricity vector.}
\end{deluxetable}

% table that includes EJ system orbital parameters
\begin{deluxetable}{c c c}[ht!]
\tabletypesize{\footnotesize}
\tablecaption{\\ Frequency Comparison \label{tab:compare}}
\tablehead{\colhead{$e_0$} & \colhead{$f_{predict}$ [$''$/yr]} & \colhead{$f_{actual}$ [$''$/yr]} }
\startdata
0 & - & - \\
0.1 & 21.22 & 21.22 \\
0.3 & 21.99 & 21.98 \\
0.5 & 23.32 & 23.32 \\
\enddata
\tablecomments{The predicted value of the eccentricity-enabled frequency  ($f_{predict}$) for each case of initial eccentricity ($e_0$) according to Equation \ref{eqn:predicted_freq} using the frequency values from Table \ref{tab:freqs}. For comparison, we list the actual frequency value ($f_{actual}$) from the Fourier analysis, which is also listed as $f_2$ in Table \ref{tab:freqs}. This prediction does not apply to the 0 initial eccentricity case.}
\end{deluxetable} 

\begin{figure}[ht!]
\centering
\includegraphics[width=\textwidth]{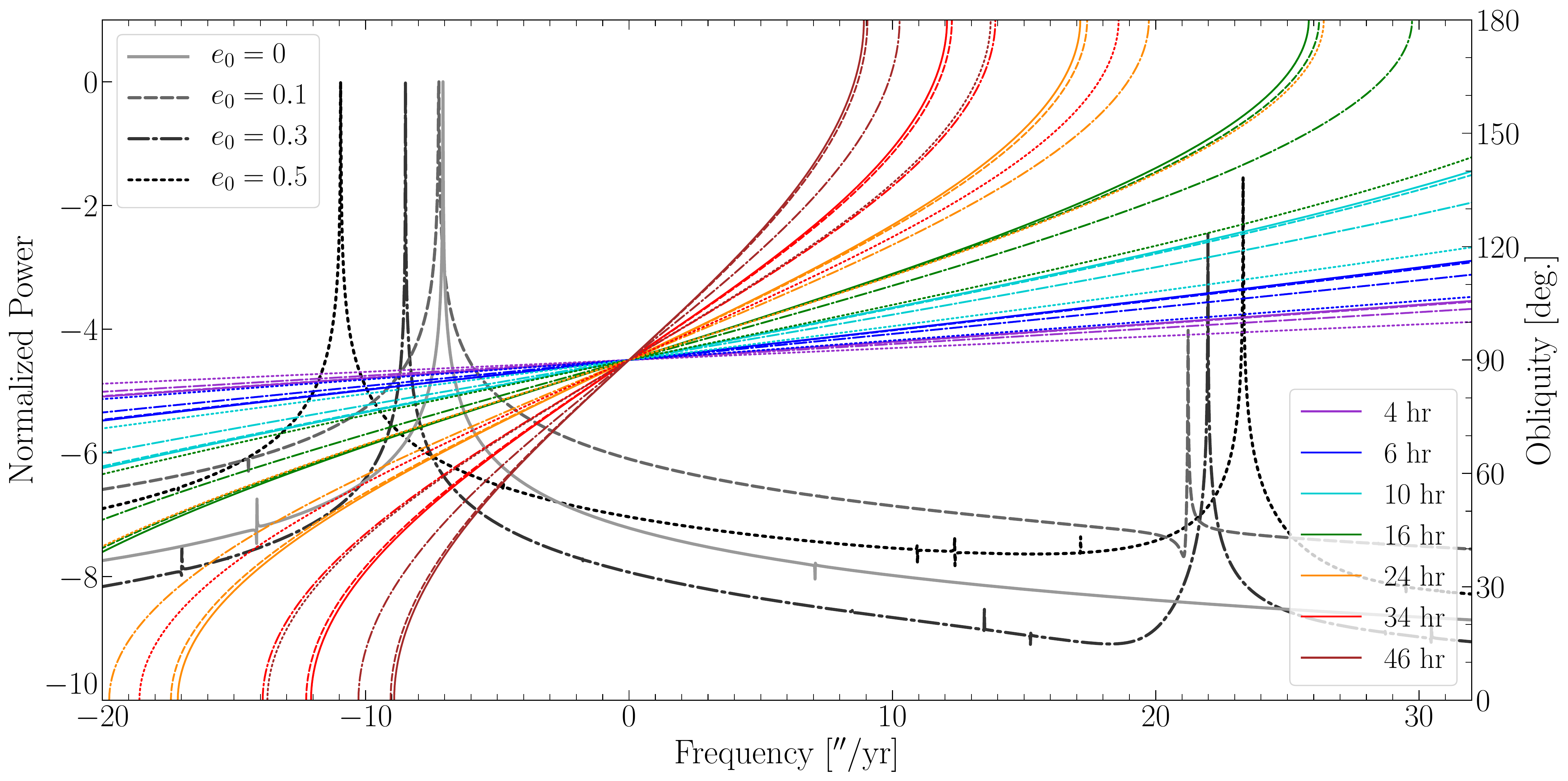}
\caption{We plot the power spectra of the orbital frequencies of Earthmoo's inclination vector for each case of initial eccentricity ($e_0$) obtained from our Frequency Modified Fourier Transform analysis. We show these spectra as the varied black lines paired with the left and bottom axes, which are normalized on a log scale. We overlay Earthmoo's range of axial precession frequencies for varied rotation periods and obliquities, computed according to Equation \ref{eqn:prec_freq}; we show these frequencies as colored lines that pair with the right and bottom axes, for each case of eccentricity (we vary the line styles the same way as the the orbital frequencies). An excitation of obliquity can occur in the case that Earthmoo's axial precession frequency becomes commensurate with an orbital frequency in a spin-orbit resonance. The positive orbital frequency peak present for the eccentric cases has the potential to enter a spin-orbit resonance with retrograde-rotating Earthmoos.}
\label{fig:spectra}
\end{figure}

Looking to Figure \ref{fig:spectra}, for each case of eccentricity, we show the orbital frequencies of the inclination vector; we overlay Earthmoo's range of axial precession frequencies according to Equation \ref{eqn:prec_freq} to allow for a visual comparison. We expect pronounced obliquity variation to occur near regions of parameter space in which these frequencies have values near one another. Inspecting the values within Table \ref{tab:freqs}, a significant positive frequency peak appears for the eccentric cases, yet is absent for the zero eccentricity case. This peak grows in amplitude for increasing values of eccentricity. Also observe that both the dominant negative peak and the positive peak shift to higher frequencies with larger values of eccentricity due to Earthmoo's orbit becoming easier to torque, which causes its ascending node to precesses faster. 

% Fourier spectra figure of orbital and rotational frequencies

We describe in Section \ref{sec:conceptual} that this positive frequency should follow a predictable relationship stated in Equation \ref{eqn:predicted_freq}. Noting that the dominant frequency for the inclination vector for each of the eccentricity cases in Table \ref{tab:freqs}, $f_1$, corresponds to $\dot{\Omega}$, the driver of nodal precession. Similarly the dominant frequency found in the analysis of the eccentricity vector in Table \ref{tab:freqs}, $g_1$, corresponds to $\dot{\varpi}$, the driver of apsidal precession. Taking the 0.3 eccentricity Earthmoo case as an example, we find that $\dot{\Omega}=f_1=-8.49$ $''$/yr and $\dot{\varpi}=g_1=6.75$ $''$/yr. According to Equation \ref{eqn:predicted_freq}, this yields $f=2\dot{\varpi}-\Omega=21.99$ $''$/yr, which nearly matches our reported value from the Fourier analysis for $f_2$ of $21.98$ $''$/yr. We compare the frequency values for the other eccentricity cases in Table \ref{tab:compare}. These cases provide good support for our proposed mechanism and show that the frequency relationship described in Equation \ref{eqn:predicted_freq} holds.

\begin{figure}[ht!]
\centering
\includegraphics[width=0.75\textwidth]{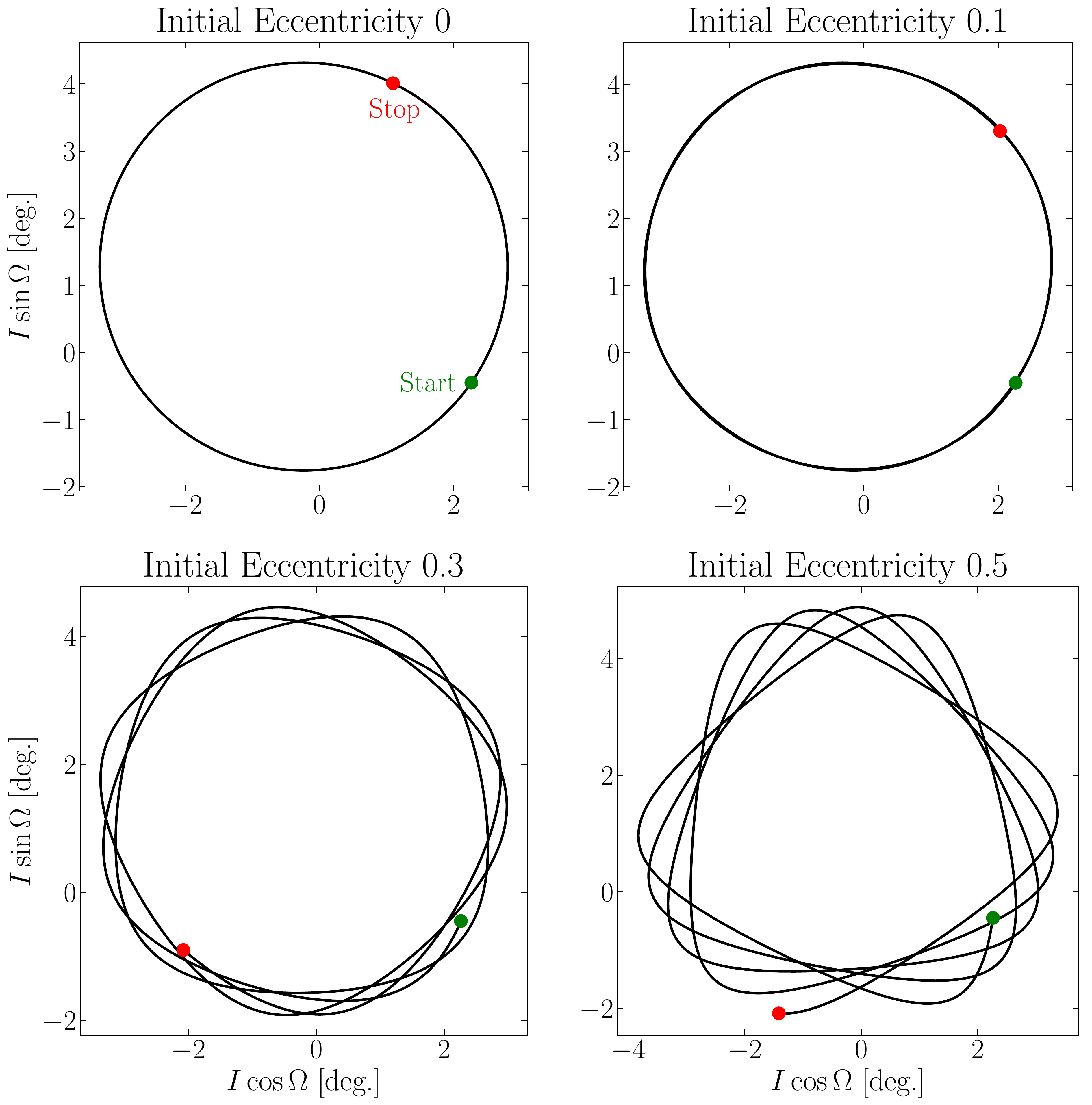}
\caption{For each case of initial eccentricity we plot the inclination vector components as [$I\sin{\Omega}$] versus [$I\cos{\Omega}$] over the course 500,000 years, sampled every 100 years. This depiction serves as a projection of the direction of Earthmoo's orbital angular momentum vector over time; the green and red dots mark the start and end of the simulation, respectively. The 0 and 0.1 eccentricity cases show Earthmoo's orbital angular momentum vector tracing out a nearly circular path as its inclination varies and its ascending node precesses. However, the 0.3 and 0.5 eccentricity cases show a path that deviates from circular, in which the orbital angular momentum vector swings outward periodically. This motion is due to the eccentricity mechanism described in Section \ref{sec:conceptual}, in which the orbital angular momentum vector undergoes a nutation in the opposing direction.}
\label{fig:pq}
\end{figure}

In addition, using the same results but only the first 500,000 years of our simulations, we generate plots of the inclination vector components as seen in Figure \ref{fig:pq}. In essence, these plots demonstrate the projection of a planet's orbital angular momentum vector as it reorients in space due to variations in the orbital inclination and the process of nodal precession. We examine these plots for each Earthmoo-Jupiter test case in order to gain additional understanding into the physical process that takes place. While the 0 and 0.1 eccentricity cases show the orbital angular momentum vector trace a neat, mostly circular path, the 0.3 and 0.5 eccentricity cases reveal more complex behavior. These higher eccentricity cases show that the position of the orbital angular momentum vector strays from circular, and swings outward periodically to produce the outer ``petals" seen with the overlap of previous tracings. This observation exemplifies the physical process described in Figures \ref{fig:nutation} and \ref{fig:precessions}, in which the additional positive frequency introduced acts to nutate the orbital angular momentum vector in the opposing direction as it undergoes nodal precession.

% pq plots for each of the eccentricity cases

\subsection{N-body Simulations} \label{sec:N-body}
The previous section showed and discussed the nature of the secular orbital eigenfrequencies in our toy Earthmoo-Jupiter system. The mechanism that we describe in Section \ref{sec:conceptual} successfully predicts a significant positive secular frequency that could participate in a spin-orbit resonance for a retrograde-rotating Earth. However, we must also determine whether this resonance is significant enough to drive large-amplitude obliquity variations important for the consideration of life. 

We demonstrate this phenomenon by simulating the aforementioned Earthmoo-Jupiter system over the course of 100 Myr, sampling at 10,000 year intervals with an integration time step of $5\%$ of Earthmoo's orbital period. Note that orbital changes of Earthmoo's inclination and ascending node throughout the simulations result in changes in its obliquity, so we expect a baseline obliquity variation of up to $\Delta \Psi = \Psi_{max} - \Psi_{min} \approx 2I_m \sim 6^\circ$ \citep{Quarles_2019}, where $I_m$ is the mutual inclination between the orbits of Earthmoo and Jupiter as stated in Section \ref{sec:initial}. We are interested in pronounced obliquity variations that arise from the proposed spin-orbit resonance, which we expect to be nestled among this baseline variation. We describe the results in the following section, with frequent reference to Table \ref{tab:freqs} and Figure \ref{fig:spectra} in order to discuss their relations.

\begin{figure}[ht!]
\centering
\includegraphics[width=\textwidth]{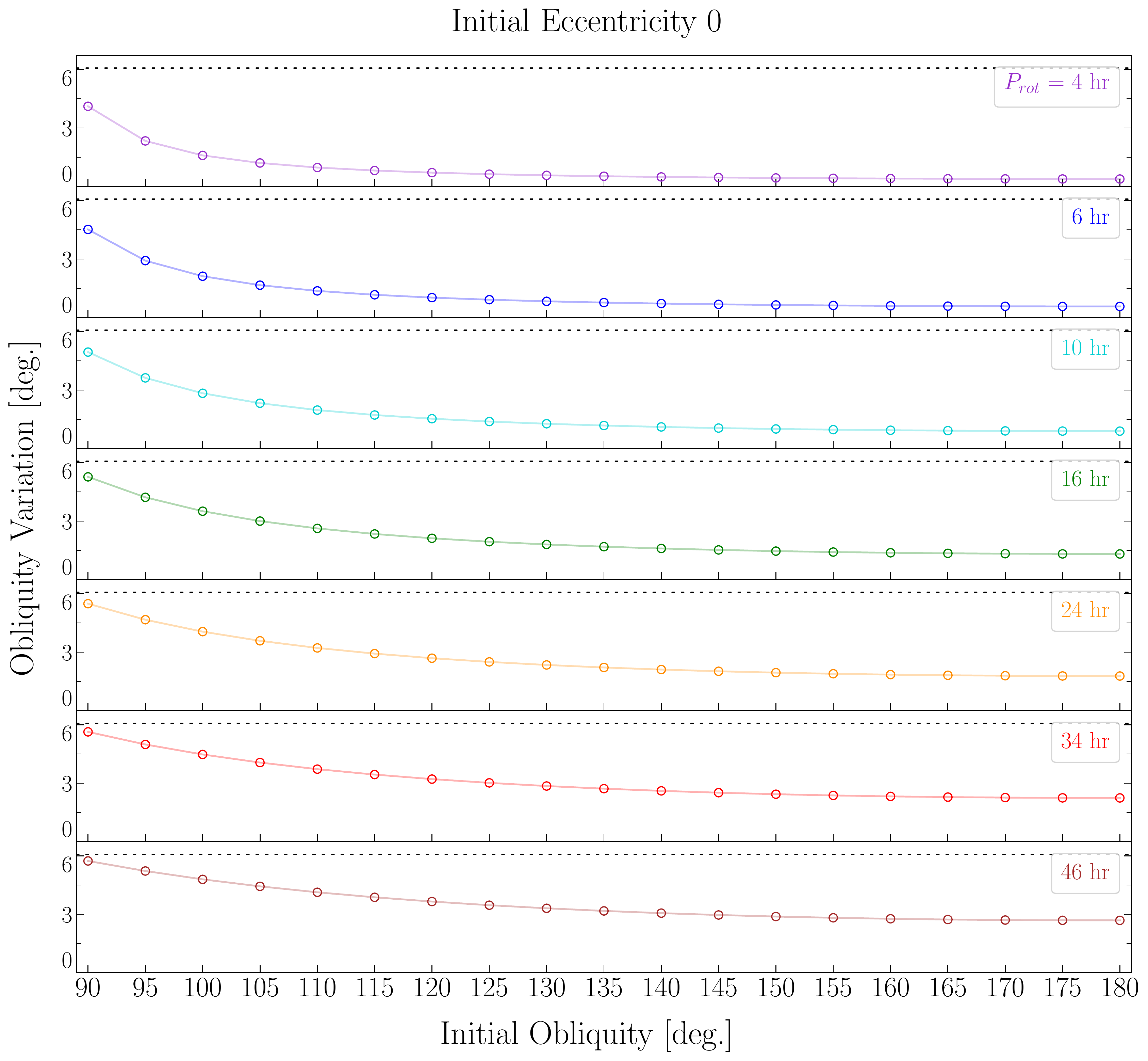}
\caption{For the Earthmoo case with an initial eccentricity of zero, we show the range of obliquity variation ($\Delta \Psi$) explored by each Earthmoo ghost planet over the course of the 100 Myr simulation as the colored markers with the vertical axis. The black dashed lines mark the threshold of the expected baseline obliquity variation due solely to orbital changes (twice the mutual inclination of Earthmoo and Jupiter). The horizontal axis corresponds to the initial obliquity we set each Earthmoo to at the start of the simulation, while each subplot groups corresponds to the different rotation periods ($P_{rot}$) of 4, 6, 10, 16, 24, 34, and 46 hours that we test. The color scheme is consistent with that of Figure \ref{fig:spectra}.}
\label{fig:0ecc}
\end{figure}

Beginning with the zero initial eccentricity case, Figure \ref{fig:0ecc} displays the range of obliquity variation explored by each Earthmoo for the tested parameter space over the course of the 100 Myr simulation. For the rotation periods and initial obliquities tested, the majority of Earthmoos experience little variation in their obliquity across the board, mostly varying $\sim$ $1^\circ$-$3^\circ$, similar to that of the present day Earth. Here we find that $100\%$ of the Earthmoos vary less than the expected $2I_m$ baseline value. However, an exception here is a pronounced increase in obliquity variation for the Earthmoos with initial obliquities set near $90^\circ$. These cases experience obliquity variations of $\sim$ $3^\circ$-$6^\circ$. Although these are still relatively small values, looking back to Figure \ref{fig:spectra}, the proximity of the axial precession frequencies of these $90^\circ$ obliquity Earthmoos to the dominant negative orbital frequency explains the pronounced variation. Looking to Table \ref{tab:freqs}, the frequency $f_1$ is the primary driver of nodal precession, clocking at -7.07 $''$/yr. Recall that for the zero eccentricity  case, there is no significant positive orbital frequency present, hence the general lack of interesting variation across the parameter space.

% Figure for the obliquity ranges of the 0 eccentricity case

\begin{figure}[ht!]
\centering
\includegraphics[width=\textwidth]{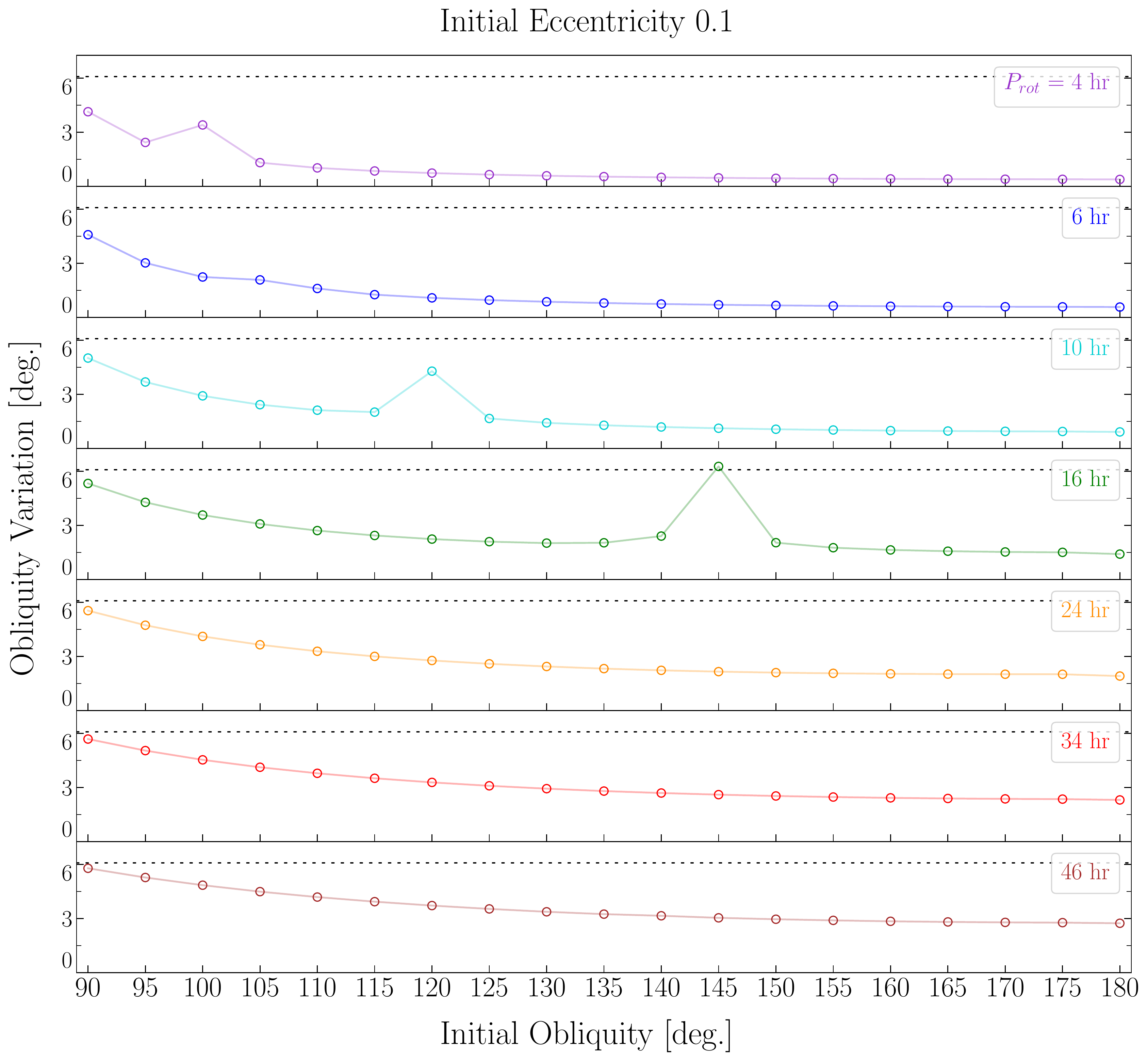}
\caption{Similar to Figure \ref{fig:0ecc}, but for the Earthmoo case with an initial eccentricity of 0.1.}
\label{fig:0.1ecc}
\end{figure}

Moving on to the 0.1 initial eccentricity case, as a similar plot, Figure \ref{fig:0.1ecc} displays the range of obliquity variation for this set of Earthmoos. Again the majority of test cases exhibit obliquity variations $\sim$ $1^\circ$-$3^\circ$ ($99.2\%$ vary less than $2I_m$), with the Earthmoos having nearly $90^\circ$ obliquities showing variations of $\sim$ $4^\circ$-$6^\circ$. Interestingly, there are a few additional test Earthmoos that stick out with pronounced obliquity variations. These include the $100^\circ$ obliquity Earthmoo in the 4 hour case varying $\sim 3^\circ$, the $120^\circ$ obliquity Earthmoo in the 10 hour case varying $\sim 4^\circ$, and the $145^\circ$ obliquity Earthmoo in the 16 hour case varying $\sim 6^\circ$. Figure \ref{fig:spectra} maps these regions of parameter space, where the Earthmoos of these combinations of rotation period and obliquity have frequencies near to the now excited positive orbital frequency peak which Table \ref{tab:freqs} marks as 21.23 $''$/yr.

% Figure for the obliquity ranges of the 0.1 eccentricity case

\begin{figure}[ht!]
\centering
\includegraphics[width=\textwidth]{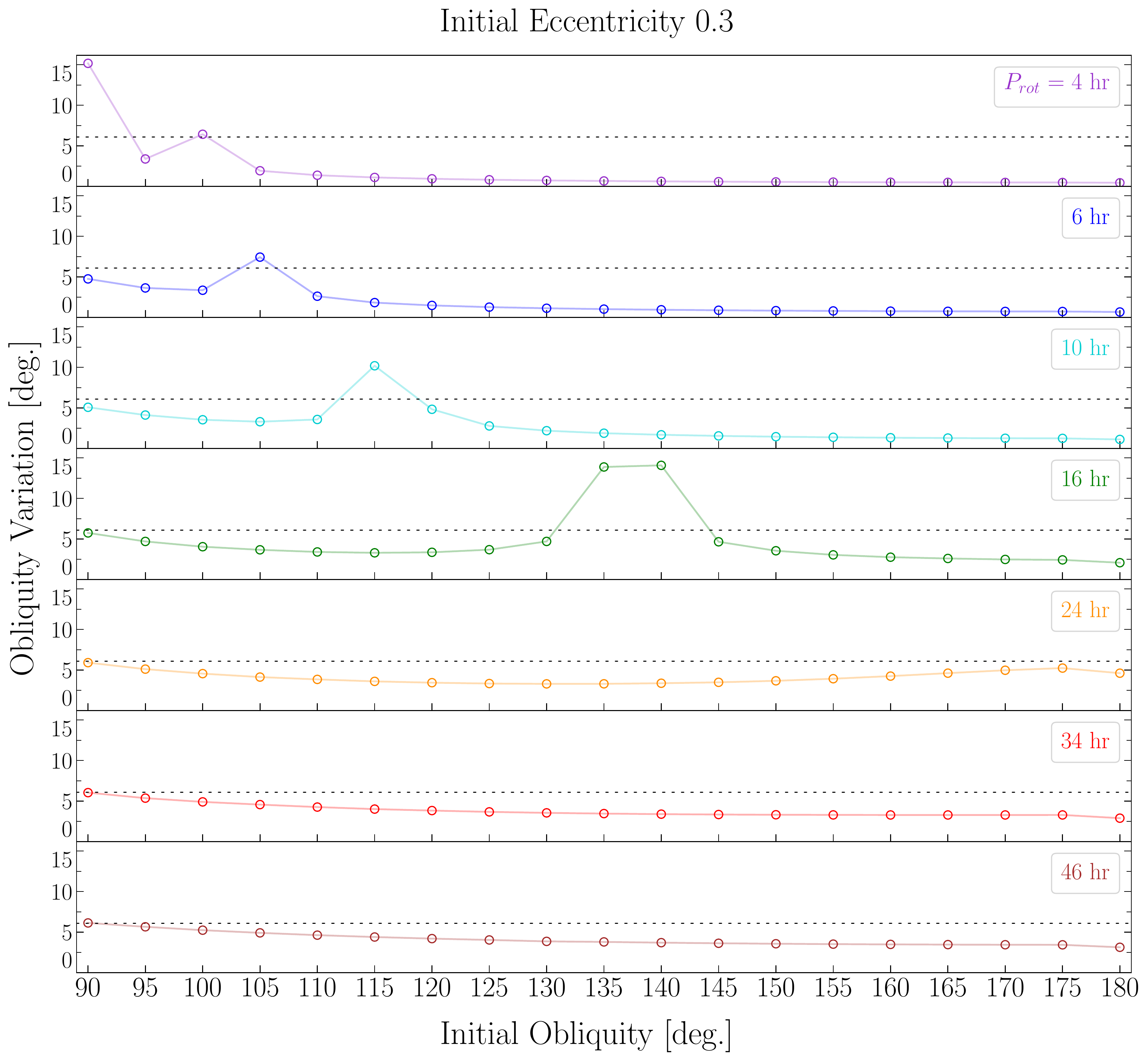}
\caption{Similar to Figure \ref{fig:0ecc}, but for the Earthmoo case with an initial eccentricity of 0.3.}
\label{fig:0.3ecc}
\end{figure}

The 0.3 initial eccentricity case follows this trend, looking to Figure \ref{fig:0.3ecc}. Most of these Earthmoo cases see obliquity variations of $\sim$ $1^\circ$-$4^\circ$ ($94.7\%$ vary less than the $2I_m$), while the $90^\circ$ obliquity Earthmoos swing $\sim$ $5^\circ$-$6^\circ$ with rotation periods 6-46 hours. The 4 hour $90^\circ$ obliquity Earthmoo in particular experienced a much larger $\sim 15^\circ$ range of variation. The 4 hour case also has an additional ``hot spot" for the Earthmoo with an initial set obliquity of $100^\circ$, varying by $\sim 6^\circ$. The 6 hour case has an Earthmoo with a pronounced $\sim 7^\circ$ variation with an initial $105^\circ$ obliquity. Additionally, the 10 hour case has Earthmoos that experience $\sim 10^\circ$ and $5^\circ$ variations for the initial obliquities $115^\circ$ and $120^\circ$, respectively. The 16 hour case sees pronounced variation centered around the $135^\circ$ and $140^\circ$ obliquity cases each with $\sim 14^\circ$ variation. The 24 hour case is not as dramatic, but sees slightly pronounced variation for its higher obliquity cases $> 155^\circ$ by a few degrees. According to Table \ref{tab:freqs}, the positive orbital frequency peak has increased in amplitude and has been shifted to a frequency of 21.98 $''$/yr. The shift of the peak appears to correlate roughly with the regions of pronounced obliquity variation as mapped by Figure \ref{fig:spectra}.

% Figure for the obliquity ranges of the 0.3 eccentricity case

\begin{figure}[ht!]
\centering
\includegraphics[width=\textwidth]{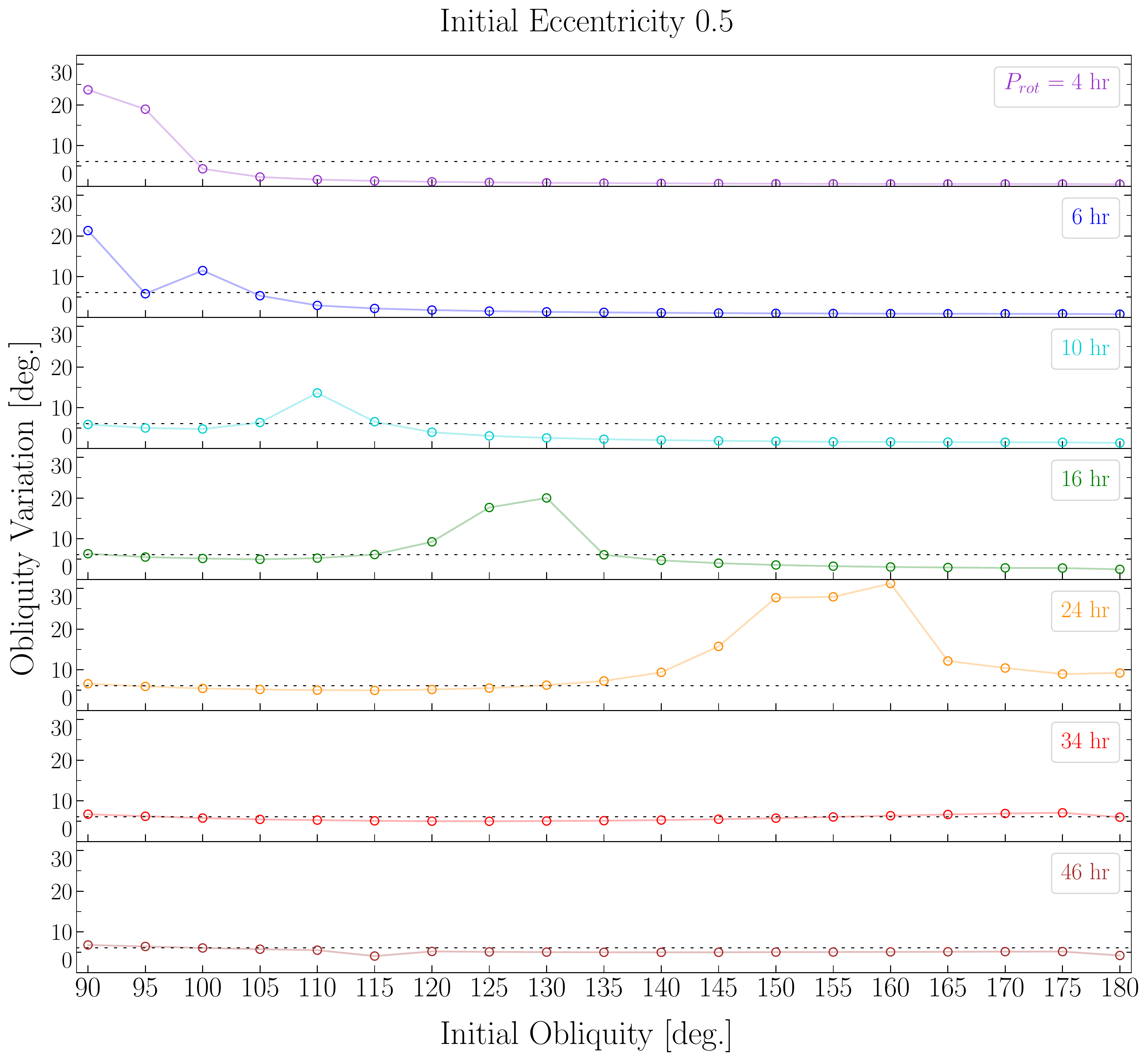}
\caption{Similar to Figure \ref{fig:0ecc}, but for the Earthmoo case with an initial eccentricity of 0.5.}
\label{fig:0.5ecc}
\end{figure}

Finally, Figure \ref{fig:0.5ecc} displays the results of the 0.5 initial eccentricity case. Most of test Earthmoo cases with rotation periods 4-16 hours only experience obliquity variations of $\sim$ $1^\circ$-$4^\circ$. The majority of Earthmoos with rotation periods 24-46 hours see consistent yet higher variations of $\sim$ $5^\circ$-$6^\circ$. Here we find that $75.9\%$ of the Earthmoos experience obliquity variations less than the expected $2I_m$ baseline value. This time, the Earthmoos with initial obliquities of $90^\circ$ don't stick out quite as much, although the 4 hour and 6 hour case Earthmoos do experience variations of $\sim 24^\circ$ and $21^\circ$, respectively. The driver of the large obliquity variation for the $90^\circ$ obliquity Earthmoos has appeared to be the fault of the negative orbital frequency peak at -10.95 $''$/yr according to Table \ref{tab:freqs}. This peak appears to have now shifted enough so that the Earthmoos with rotation periods $>$ 6 hours evade a resonance with it. Notable test cases with pronounced variation include the four hour $95^\circ$ obliquity Earthmoo that varies $\sim 19^\circ$, the 6 hour $100^\circ$ obliquity Earthmoo that varies $\sim 11^\circ$, the 10 hour $110^\circ$ obliquity Earthmoo that varies $\sim 14^\circ$, the 16 hour $125^\circ$ and $130^\circ$ obliquity Earthmoos that vary about $18^\circ$ and $20^\circ$, respectively, and the 24 hour batch of Earthmoos with initial obliquities ranging from $145^\circ$ to $160^\circ$ that vary from $16^\circ$ to $31^\circ$. The positive frequency orbital peak has again increased in amplitude due to the greater value of eccentricity, and is roughly 23.32 $''$/yr according to Table \ref{tab:freqs}. Figure \ref{fig:spectra} once more largely maps out the pattern of variation.

% Figure for the obliquity ranges of the 0.5 eccentricity case

\subsection{Semi-Analytical Secular Model} \label{sec:semi-analytical}
Although we are confident in our methodology and the accuracy of our results, running simulations with \texttt{SMERCURY} is computationally expensive and limits the parameter space that we can explore. We choose to complement our results by examining a broader parameter space and employ the less robust but useful semi-analytical secular spin evolution approach as featured in \citet{Quarles_2019}. This approach applies a secular time-dependent Hamiltonian to the orbital integration results computed by \texttt{SMERCURY} in order to compute the spin evolution of a planet. We use the same initial conditions as our other simulations (according to Table \ref{tab:orbital}), and apply this model to the 100 Myr, sampled every 100 years, orbital evolution results as featured in Figures \ref{fig:spectra} and \ref{fig:pq}. However, we now explore initial obliquities for Earthmoo from $90^\circ$ to $180^\circ$ in $1^\circ$ increments while we sweep the values of the precession constant (relates to rotation period) from 0 $''$/yr to 120 $''$/yr in 1 $''$/yr increments. Here the initial precession angle is $348^\circ$.

\begin{figure}[ht!]
\centering
\includegraphics[width=\textwidth]{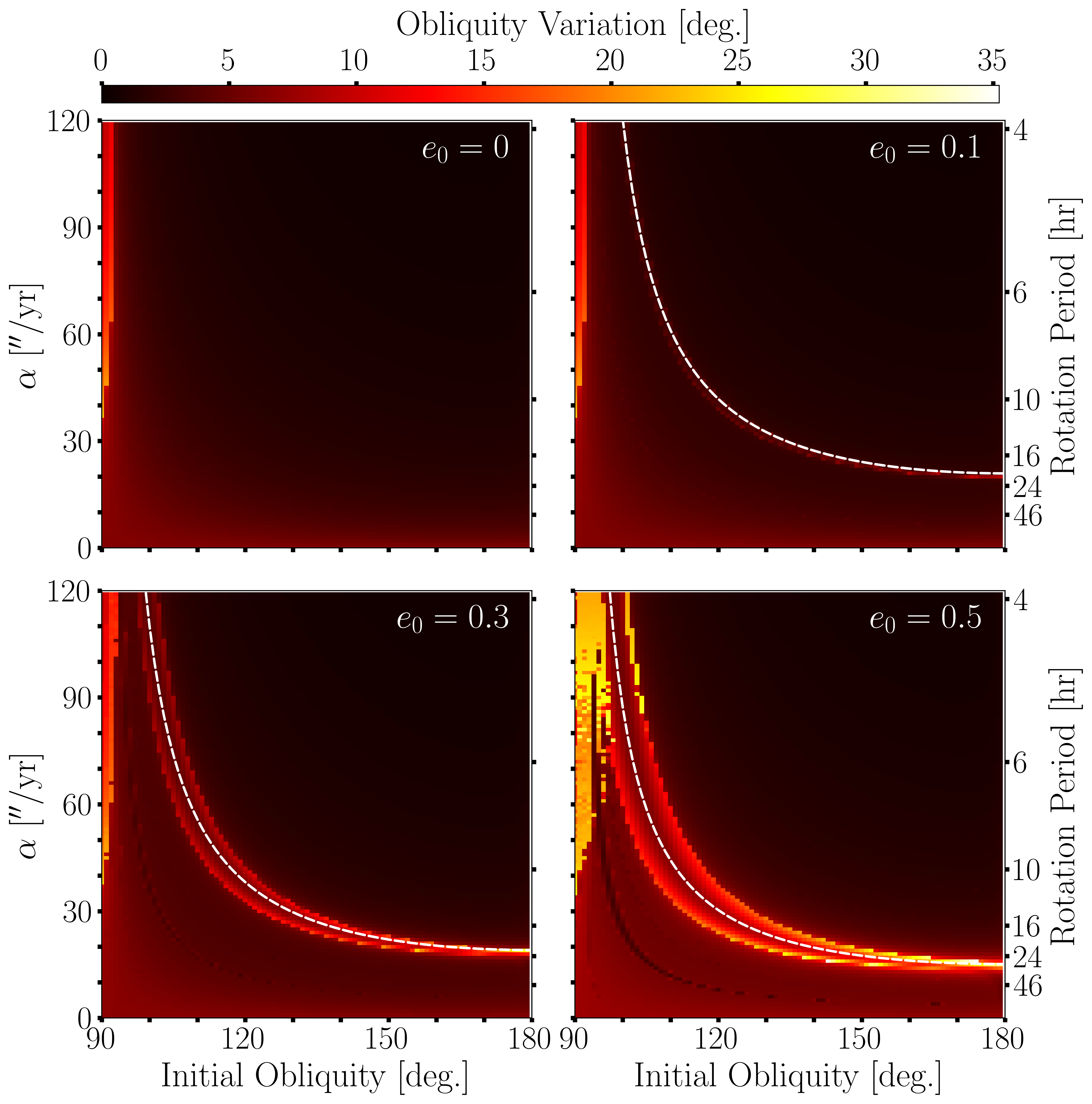}
\caption{For each case of initial eccentricity ($e_0$) we assign, we show the obliquity variation ($\Delta \Psi$) each Earthmoo experienced over the course of the 100 Myr simulation for a range of values of the initial obliquity and precession constant, $\alpha$ (related to the rotation period). The values of obliquity variation are mapped to the color bar shown at the top of the plot, ranging from zero to the maximum recorded value of $35.24^\circ$ (from the $e_0 = 0.5$ case) according to the semi-analytical approach described in Section \ref{sec:semi-analytical}. The dashed white lines for the 0.1, 0.3, and 0.5 eccentricity cases mark the predicted locations of the spin-orbit resonance.}
\label{fig:Billy}
\end{figure}

Looking to Figure \ref{fig:Billy}, we display the range of obliquity variation ($\Delta \Psi$) each Earthmoo experienced for each case of initial eccentricity. Since we were able to explore a much finer grid of parameter space using this approach, we present our results in different way that more easily highlights the trends. These results are in good agreement with those discussed in Section \ref{sec:N-body}. The majority of Earthmoos are quite stable and only vary $1^\circ$-$2^\circ$, while the cases in which Earthmoo had an initial obliquity near $90^\circ$ similarly resulted in heightened obliquity variation. The test Earthmoos experienced pronounced obliquity variation near the predicted locations of the spin-orbit resonance. In addition, there is a splitting trend that distinguishes two regimes of variation that follow the predicted resonance location which become more evident for higher values of eccentricity. Figure 12 of \citet{Quarles_2020} showcases a similar splitting trend, in which the authors explain that it is a consequence as a result of the difference between the initial precession angle and the initial longitude of ascending node ($\phi_0 - \Omega_0$) having a large angular separation from the phase of the resonant orbital frequency (shown as $\gamma_i$ in Table \ref{tab:freqs}). We again find that the magnitude of these variations generally increase with eccentricity (due to increasing amplitudes of the orbital frequencies), where we report a maximum variation of $35.24^\circ$ for the Earthmoo with an initial eccentricity of 0.5 with an initial obliquity of $174^\circ$ and a precession constant of 16 $''$/yr.

% Billy plot which uses secular approach but captures all of the results for easy viewing

\subsection{A Closer Look at the Resonance} \label{sec:libration}
Taking the results of our N-body simulations, we can gain additional insight into the nature of this eccentricity-enabled spin-orbit resonance through further analysis. In Section \ref{sec:conceptual} we discussed that the condition for this resonance was that $2 \dot{\varpi} - \dot{\Omega} - \dot{\phi} = 0$. The corresponding resonant angle, $\theta$, is therefore

\begin{equation}
    \theta = 2 \varpi - \Omega - \phi
    \label{eqn:resonant_angle}
\end{equation}

\begin{figure}[ht!]
\centering
\includegraphics[width=\textwidth]{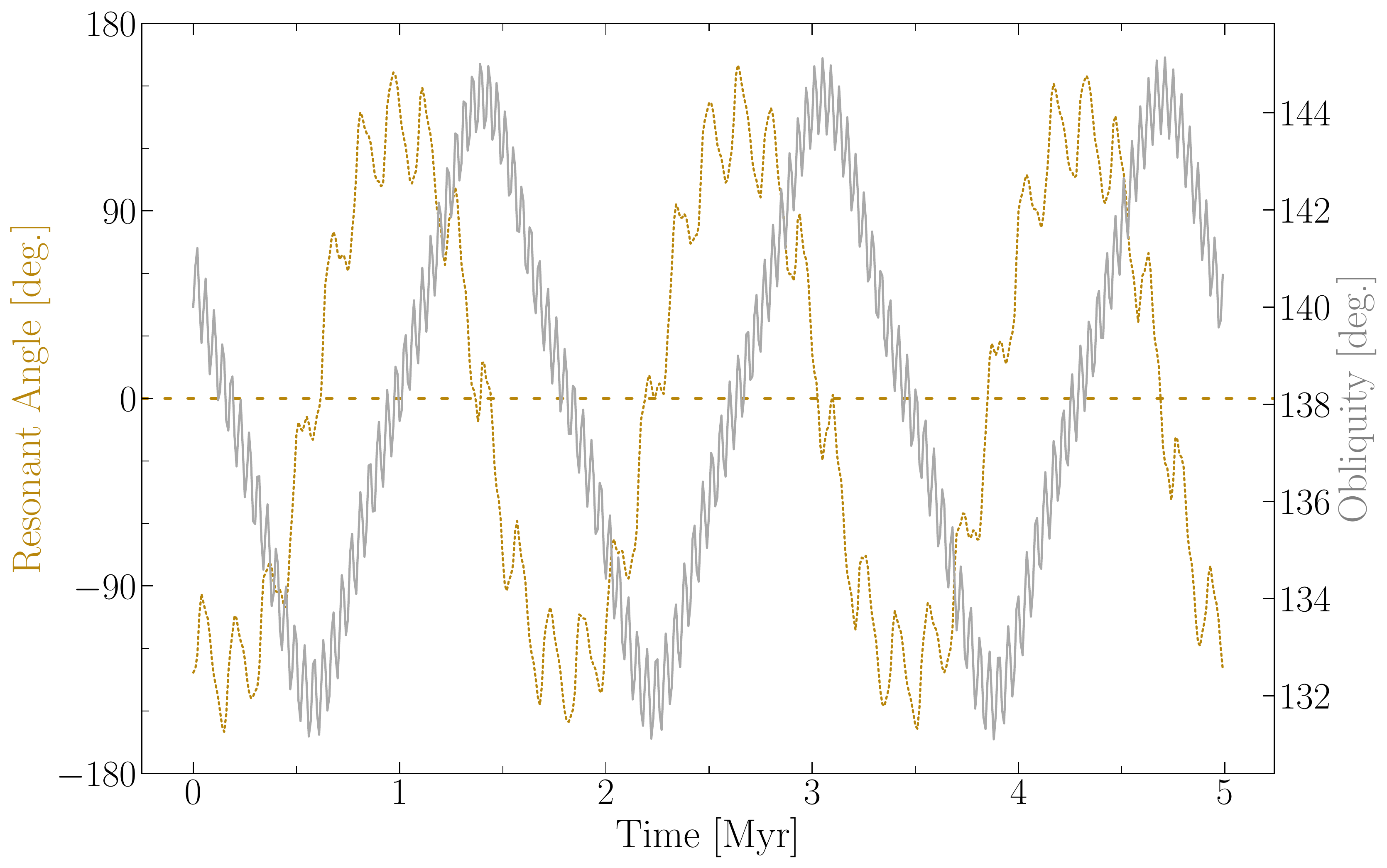}
\caption{The first 5 Myr of the evolution of the resonant angle and obliquity of the Earthmoo on an orbit with an initial eccentricity of 0.3 with a 16-hour rotation period and an initial obliquity of $140^\circ$. We display the resonant angle from Equation \ref{eqn:resonant_angle} as the dotted gold line that pairs with the left vertical axis. The dashed horizontal gold line marks the $0^\circ$ location for the resonant angle. We show the Earthmoo's obliquity as the solid silver line that pairs with the right vertical axis.}
\label{fig:libration}
\end{figure}

\noindent \citet{Saillenfest_2019} provides an analytical theory that identifies this angle as a first-order term that obeys the ``Colombo's Top" Hamiltonian. Here, we use our numerical simulations to verify that $\theta$ is indeed a valid resonant angle, and that it is connected to the observed obliquity variations. We select the Earthmoo with a 16-hour rotation period and an initial obliquity of $140^\circ$ from the 0.3 initial eccentricity case (featured in Figure \ref{fig:0.3ecc}) to serve as an example. We plot the Earthmoo's resonant angle and obliquity as functions of time for the first 5 Myr of the simulation in Figure \ref{fig:libration}. This plot is revealing, and shows that $\theta$ is indeed librating about zero, as one would expect for a valid resonant angle. Furthermore, the variations in $\theta$ lead the variations in obliquity by $90^\circ$, implying a direct dynamical connection between these two angles.

The behavior shown in Figure \ref{fig:libration} also highlights the importance of initial conditions. Previous work found that the choice of the initial precession angle can influence the obliquity evolution in the sense that it affects the phasing of frequencies involved in the spin-orbit resonance, which in turn moderates the strength of the resonant interactions \citep{Lissauer_2012, Barnes_2016, Quarles_2020}. The example Earthmoo in Figure \ref{fig:libration} had an initial precession angle of $348.74^\circ$, in which the wide libration of its resonant angle and subsequent large obliquity variations was likely determined by its initial combination of $\varpi$, $\Omega$, and $\phi$. Therefore we conclude that these angles play a role in controlling the range of affected obliquity configurations near the resonance. Future work could explore this phenomenon.

\section{Conclusion} \label{sec:conclusion}
We investigate a mechanism in which orbital eccentricity acts to influence the obliquity evolution of retrograde-rotating planets. This mechanism can lead to pronounced obliquity variations in the event of a spin-orbit resonance, when a retrograde rotator's axial precession frequency becomes commensurate with an eccentricity-enabled orbital frequency. We demonstrate this phenomenon by performing numerical simulations of a toy system consisting of the Sun, a moonless Earth we call Earthmoo, and Jupiter. 

We find that the predicted orbital frequency is indeed present for the cases in which Earthmoo is eccentric; this orbital frequency obeys the relationship we describe in Equation \ref{eqn:predicted_freq}, while it shifts to higher values and grows in amplitude for greater values of eccentricity. Over the course of the 100 Myr rigid body simulations we find that $92.5\%$ of the Earthmoos that we test are relatively obliquity stable and experienced variations less than what we expect from orbital variations alone. However, for regions of parameter space near the predicted spin-orbit resonance locations, Earthmoo experienced pronounced obliquity variations of order $10^\circ$-$30^\circ$. Generally, greater values of eccentricity led to an increase in the fraction of parameter space occupied by solutions with large obliquity variations, in which these variations also increased with increasing eccentricity.

The magnitude of these obliquity variations over this timescale is likely significant in the consideration of habitability. We do not employ a climate model in this work similar to \citet{Armstrong_2014}, \citet{Deitrick_2018b}, and others, but others have shown that large swings in a planet's obliquity could be the deciding factor between conditions that are detrimental to life, or foster it. Although the system that we consider here is purely fictional, we show the importance of the consideration of orbital eccentricity for retrograde-rotating planets and suggest that this effect may be important in future efforts to characterize planetary habitability. Retrograde rotation stabilizes obliquity under most circumstances, but the obliquity evolution of a planet is ultimately at the mercy of its system's orbital architecture. Retrograde rotators on eccentric orbits can experience high obliquity variability. 

Ideally, the knowledge gained by this study and the numerous other obliquity studies could be applied to real exoplanet systems. However, to date, we have not yet been able to directly measure the obliquity of an exoplanet. There are several methodologies that have been proposed to do so, including: constraining obliquity by measuring planetary oblateness and rotation period from transit photometry \citep{Barnes_2003}, inferring obliquity from seasonal differences in infrared light curves \citep{Gaidos_2004}, obtaining obliquity by looking for geometric effects of directly imaged exoplanets \citep{Kawahara_2016}, and others. It is only a matter of time before we gain this capability with future technological advances, where we can then apply our knowledge and make judgements on the habitability of candidate exoplanets based on their obliquity evolution.

\section{Acknowledgments} \label{sec:ack}
The authors acknowledge support from the NASA Exobiology program, grant \#NNX14AK31G. This research was supported in part through research cyberinfrastructure resources and services provided by the Partnership for an Advanced Computing Environment (PACE) at the Georgia Institute of Technology. We thank William J. Miller for his code to produce the nodal precession animation featured in Figure \ref{fig:ani_nodal}.

\section{Orcid iDs} \label{sec:orcid}
\noindent Steven M. Kreyche \orcidicon{0000-0002-7274-758X} \url{https://orcid.org/0000-0002-7274-758X}
\\
Jason W. Barnes \orcidicon{0000-0002-7755-3530} \url{https://orcid.org/0000-0002-7755-3530}
\\
Billy L. Quarles \orcidicon{0000-0002-9644-8330} \url{https://orcid.org/0000-0002-9644-8330}
\\
Jack J. Lissauer \orcidicon{0000-0001-6513-1659} \url{https://orcid.org/0000-0001-6513-1659}
\\
Matthew M. Hedman \orcidicon{0000-0002-8592-0812} \url{https://orcid.org/0000-0002-8592-0812}

\bibliography{Kreyche_2020}

\begin{thebibliography}{}
\expandafter\ifx\csname natexlab\endcsname\relax\def\natexlab#1{#1}\fi
\providecommand{\url}[1]{\href{#1}{#1}}

\bibitem[{Armstrong {et~al.}(2014)Armstrong, Barnes, Domagal-Goldman, Breiner,
  Quinn, \& Meadows}]{Armstrong_2014}
Armstrong, J., Barnes, R., Domagal-Goldman, S., {et~al.} 2014, Astrobiology,
  14, 277, pMID: 24611714.
\newblock \url{https://doi.org/10.1089/ast.2013.1129}

\bibitem[{Barnes \& Fortney(2003)}]{Barnes_2003}
Barnes, J.~W., \& Fortney, J.~J. 2003, The Astrophysical Journal, 588, 545.
\newblock \url{https://doi.org/10.1086\%2F373893}

\bibitem[{Barnes {et~al.}(2016)Barnes, Quarles, Lissauer, Chambers, \&
  Hedman}]{Barnes_2016}
Barnes, J.~W., Quarles, B., Lissauer, J.~J., Chambers, J., \& Hedman, M.~M.
  2016, Astrobiology, 16, 487, pMID: 27328026.
\newblock \url{https://doi.org/10.1089/ast.2015.1427}

\bibitem[{{Bolmont} {et~al.}(2016){Bolmont}, {Libert}, {Leconte}, \&
  {Selsis}}]{Bolmont_2016}
{Bolmont}, E., {Libert}, A.-S., {Leconte}, J., \& {Selsis}, F. 2016, A\&A, 591,
  A106.
\newblock \url{https://doi.org/10.1051/0004-6361/201628073}

\bibitem[{Chambers(1999)}]{Chambers_1999}
Chambers, J.~E. 1999, Monthly Notices of the Royal Astronomical Society, 304,
  793.
\newblock \url{https://doi.org/10.1046/j.1365-8711.1999.02379.x}

\bibitem[{Colose {et~al.}(2019)Colose, Genio, \& Way}]{Colose_2019}
Colose, C.~M., Genio, A. D.~D., \& Way, M.~J. 2019, The Astrophysical Journal,
  884, 138.
\newblock \url{https://doi.org/10.3847%2F1538-4357%2Fab4131}

\bibitem[{{Deitrick} {et~al.}(2018a){Deitrick}, {Barnes}, {Quinn}, {Armstrong},
  {Charnay}, \& {Wilhelm}}]{Deitrick_2018a}
{Deitrick}, R., {Barnes}, R., {Quinn}, T.~R., {et~al.} 2018a, The Astronomical
  Journal, 155, 60.
\newblock \url{https://doi.org/10.3847%2F1538-3881%2Faaa301}

\bibitem[{{Deitrick} {et~al.}(2018b){Deitrick}, {Barnes}, {Bitz}, {Fleming},
  {Charnay}, {Meadows}, {Wilhelm}, {Armstrong}, \& {Quinn}}]{Deitrick_2018b}
{Deitrick}, R., {Barnes}, R., {Bitz}, C., {et~al.} 2018b, The Astronomical
  Journal, 155, 266.
\newblock \url{https://doi.org/10.3847%2F1538-3881%2Faac214}

\bibitem[{{Dones} \& {Tremaine}(1993)}]{Dones_1993}
{Dones}, L., \& {Tremaine}, S. 1993, \icarus, 103, 67

\bibitem[{Dong {et~al.}(2019)Dong, Huang, \& Lingam}]{Dong_2019}
Dong, C., Huang, Z., \& Lingam, M. 2019, The Astrophysical Journal, 882, L16.
\newblock \url{https://doi.org/10.3847%2F2041-8213%2Fab372c}

\bibitem[{Dressing {et~al.}(2010)Dressing, Spiegel, Scharf, Menou, \&
  Raymond}]{Dressing_2010}
Dressing, C.~D., Spiegel, D.~S., Scharf, C.~A., Menou, K., \& Raymond, S.~N.
  2010, The Astrophysical Journal, 721, 1295.
\newblock \url{https://doi.org/10.1088%2F0004-637x%2F721%2F2%2F1295}

\bibitem[{{Forget} {et~al.}(2013){Forget}, {Wordsworth}, {Millour},
  {Madeleine}, {Kerber}, {Leconte}, {Marcq}, \& {Haberle}}]{Forget_2013}
{Forget}, F., {Wordsworth}, R., {Millour}, E., {et~al.} 2013, \icarus, 222, 81

\bibitem[{Gaidos \& Williams(2004)}]{Gaidos_2004}
Gaidos, E., \& Williams, D. 2004, New Astronomy, 10, 67 .
\newblock \url{https://doi.org/10.1016/j.newast.2004.04.009}

\bibitem[{Guendelman \& Kaspi(2019)}]{Guendelman_2019}
Guendelman, I., \& Kaspi, Y. 2019, The Astrophysical Journal, 881, 67.
\newblock \url{https://doi.org/10.3847%2F1538-4357%2Fab2a06}

\bibitem[{Head {et~al.}(2004)Head, Mustard, Kreslavsky, Milliken, \&
  Marchant}]{Head_2004}
Head, J., Mustard, J., Kreslavsky, M., Milliken, R., \& Marchant, D. 2004,
  Nature, 426, 797

\bibitem[{{Head} {et~al.}(2005){Head}, {Neukum}, {Jaumann}, {Hiesinger},
  {Hauber}, {Carr}, {Masson}, {Foing}, {Hoffmann}, {Kreslavsky}, {Werner},
  {Milkovich}, {van Gasselt}, \& {HRSC Co-Investigator Team}}]{Head_2005}
{Head}, J.~W., {Neukum}, G., {Jaumann}, R., {et~al.} 2005, \nat, 434, 346

\bibitem[{Hoffman {et~al.}(1998)Hoffman, Kaufman, Halverson, \&
  Schrag}]{Hoffman_1998}
Hoffman, P.~F., Kaufman, A.~J., Halverson, G.~P., \& Schrag, D.~P. 1998,
  Science, 281, 13421346.
\newblock \url{https://science.sciencemag.org/content/281/5381/1342}

\bibitem[{Hubbard(1984)}]{Hubbard_1984}
Hubbard, W.~B. 1984, Planetary Interiors (New York, Van Nostrand Reinhold Co.)

\bibitem[{Kane \& Torres(2017)}]{Kane_2017}
Kane, S.~R., \& Torres, S.~M. 2017, The Astronomical Journal, 154, 204.
\newblock \url{https://doi.org/10.3847%2F1538-3881%2Faa8fce}

\bibitem[{Kang(2019)}]{Kang_2019}
Kang, W. 2019, The Astrophysical Journal, 876, L1.
\newblock \url{https://doi.org/10.3847%2F2041-8213%2Fab18a8}

\bibitem[{Kawahara(2016)}]{Kawahara_2016}
Kawahara, H. 2016, The Astrophysical Journal, 822, 112.
\newblock \url{https://doi.org/10.3847%2F0004-637x%2F822%2F2%2F112}

\bibitem[{Kilic {et~al.}(2018)Kilic, Lunkeit, Raible, \& Stocker}]{Kilic_2018}
Kilic, C., Lunkeit, F., Raible, C.~C., \& Stocker, T.~F. 2018, The
  Astrophysical Journal, 864, 106.
\newblock \url{https://doi.org/10.3847%2F1538-4357%2Faad5eb}

\bibitem[{Kilic {et~al.}(2017)Kilic, Raible, \& Stocker}]{Kilic_2017}
Kilic, C., Raible, C.~C., \& Stocker, T.~F. 2017, The Astrophysical Journal,
  844, 147.
\newblock \url{https://doi.org/10.3847%2F1538-4357%2Faa7a03}

\bibitem[{{Laskar} {et~al.}(1993a){Laskar}, {Joutel}, \&
  {Boudin}}]{Laskar_1993a}
{Laskar}, J., {Joutel}, F., \& {Boudin}, F. 1993a, \aap, 270, 522.
\newblock \url{https://ui.adsabs.harvard.edu/abs/1993A&A...270..522L}

\bibitem[{Laskar \& Robutel(1993b)}]{Laskar_1993b}
Laskar, J., \& Robutel, P. 1993b, Nature, 361, 608.
\newblock \url{https://doi.org/10.1038/361608a0}

\bibitem[{Li \& Batygin(2014)}]{Li_2014}
Li, G., \& Batygin, K. 2014, The Astrophysical Journal, 790, 69.
\newblock \url{https://doi.org/10.1088%2F0004-637x%2F790%2F1%2F69}

\bibitem[{Lissauer {et~al.}(2012)Lissauer, Barnes, \& Chambers}]{Lissauer_2012}
Lissauer, J.~J., Barnes, J.~W., \& Chambers, J.~E. 2012, Icarus, 217, 77 .
\newblock \url{https://doi.org/10.1016/j.icarus.2011.10.013}

\bibitem[{Lissauer {et~al.}(1997)Lissauer, Berman, Greenzweig, \&
  Kary}]{Lissauer_1997}
Lissauer, J.~J., Berman, A.~F., Greenzweig, Y., \& Kary, D.~M. 1997, Icarus,
  127, 65

\bibitem[{Lissauer \& Kary(1991)}]{Lissauer_1991}
Lissauer, J.~J., \& Kary, D.~M. 1991, Icarus, 94, 126 .
\newblock
  \url{http://www.sciencedirect.com/science/article/pii/001910359190145J}

\bibitem[{M{\'{e}}ndez \& Rivera-Valent{\'{\i}}n(2017)}]{Mendez_2017}
M{\'{e}}ndez, A., \& Rivera-Valent{\'{\i}}n, E.~G. 2017, The Astrophysical
  Journal, 837, L1.
\newblock \url{https://doi.org/10.3847%2F2041-8213%2Faa5f13}

\bibitem[{Miguel \& Brunini(2010)}]{Miguel_2010}
Miguel, Y., \& Brunini, A. 2010, Monthly Notices of the Royal Astronomical
  Society, 406, 1935.
\newblock \url{https://doi.org/10.1111/j.1365-2966.2010.16804.x}

\bibitem[{Milankovi\'c(1998)}]{Milankovic_1998}
Milankovi\'c, M. 1998, Canon of Insolation and the Ice-Age Problem
  [translation] (Zavod za Udžbenike i Nastavna Sredstva)

\bibitem[{Millholland \& Batygin(2019)}]{Millholland_2019}
Millholland, S., \& Batygin, K. 2019, The Astrophysical Journal, 876, 119.
\newblock \url{https://doi.org/10.3847%2F1538-4357%2Fab19be}

\bibitem[{Muller \& MacDonald(1995)}]{Muller_1995}
Muller, R.~A., \& MacDonald, G.~J. 1995, Nature, 377, 107

\bibitem[{Murray \& Dermott(1999)}]{Murray_2000}
Murray, C.~D., \& Dermott, S.~F. 1999, Solar System Dynamics (Cambridge
  University Press)

\bibitem[{{Neron de Surgy} \& {Laskar}(1997)}]{Surgy_1997}
{Neron de Surgy}, O., \& {Laskar}, J. 1997, \aap, 318, 975

\bibitem[{{Olson} {et~al.}(2019){Olson}, {Jansen}, \& {Abbot}}]{Olson_2019}
{Olson}, S.~L., {Jansen}, M., \& {Abbot}, D.~S. 2019, arXiv e-prints,
  arXiv:1909.02928

\bibitem[{Quarles {et~al.}(2020)Quarles, Barnes, Lissauer, \&
  Chambers}]{Quarles_2020}
Quarles, B., Barnes, J.~W., Lissauer, J.~J., \& Chambers, J. 2020,
  Astrobiology, 20, 73, pMID: 31613645.
\newblock \url{https://doi.org/10.1089/ast.2018.1932}

\bibitem[{Quarles {et~al.}(2019)Quarles, Li, \& Lissauer}]{Quarles_2019}
Quarles, B., Li, G., \& Lissauer, J.~J. 2019, The Astrophysical Journal, 886,
  56.
\newblock \url{https://doi.org/10.3847%2F1538-4357%2Fab46b5}

\bibitem[{{Saillenfest} {et~al.}(2019){Saillenfest}, {Laskar}, \&
  {Bou\'e}}]{Saillenfest_2019}
{Saillenfest}, M., {Laskar}, J., \& {Bou\'e}, G. 2019, A\&A, 623, A4.
\newblock \url{https://doi.org/10.1051/0004-6361/201834344}

\bibitem[{{\v Sidlichovsk\'y} \& {Nesvorn\'y}(1996)}]{Sidlichovsky_1996}
{\v Sidlichovsk\'y}, M., \& {Nesvorn\'y}, D. 1996, Celestrial Mechanics and
  Dynamical Astronomy, 65, 137.
\newblock \url{https://doi.org/10.1007/BF00048443}

\bibitem[{Spiegl {et~al.}(2015)Spiegl, Paeth, \& Frimmel}]{Spiegl_2015}
Spiegl, T., Paeth, H., \& Frimmel, H. 2015, Earth and Planetary Science
  Letters, 415, 100.
\newblock \url{https://doi.org/10.1016/j.epsl.2015.01.035}

\bibitem[{van~den Heuvel(1966)}]{Heuvel_1965}
van~den Heuvel, E. P.~J. 1966, Geophysical Journal International, 11, 323.
\newblock \url{https://doi.org/10.1111/j.1365-246X.1966.tb03086.x}

\bibitem[{Way \& Georgakarakos(2017)}]{Way_2017}
Way, M.~J., \& Georgakarakos, N. 2017, The Astrophysical Journal, 835, L1.
\newblock \url{https://doi.org/10.3847%2F2041-8213%2F835%2F1%2Fl1}

\bibitem[{Williams \& Kasting(1997)}]{Williams_1997}
Williams, D.~M., \& Kasting, J.~F. 1997, Icarus, 129, 254.
\newblock \url{https://doi.org/10.1006/icar.1997.5759}

\bibitem[{Williams \& Pollard(2002)}]{Williams_2002}
Williams, D.~M., \& Pollard, D. 2002, International Journal of Astrobiology, 1,
  61.
\newblock \url{https://doi.org/10.1017/S1473550402001064}

\bibitem[{Williams \& Pollard(2003)}]{Williams_2003}
---. 2003, International Journal of Astrobiology, 2, 1.
\newblock \url{https://doi.org/10.1017/S1473550403001356}

\end{thebibliography}



\end{document}